\documentclass[useAMS,usenatbib]{mnras}
\usepackage{graphicx}
\usepackage{epstopdf}
\usepackage{amssymb}
\usepackage{amsmath}
\usepackage{multirow}
\usepackage{diagbox}
\usepackage[normalem]{ulem}

\usepackage[labelformat=empty,labelsep=none]{subcaption}
\newcommand{\ali}{\textrm{A(Li)}}
\newcommand{\ax}{\textrm{A}(\mathrm{X})}

\newcommand{\teff}{T_{\rm{eff}}}
\newcommand{\logg}{\log g}
\newcommand{\met}{[\rm{Fe}/\rm{H}]}
\newcommand{\Dn}{\Delta \rm{NLTE}}
\newcommand{\dn}{\delta \rm{NLTE}}

\usepackage[dvipsnames]{xcolor}
\newcommand{\bb}[1]{#1}

\newcommand{\msun}{\rm{M_{\odot}}}

\newcommand{\rewlte}{\textrm{REW}_{\rm{LTE}}}
\newcommand{\rewnlte}{\textrm{REW}_{\rm{NLTE}}}

\newcommand{\mlt}{\alpha_{\textrm MLT}}
\newcommand{\vmic}{v_{\textrm{mic}}}

\newcommand{\kms}{\textrm{km\,s$^{-1}$}}
\newcommand{\blue}{\textsc{blue}}
\newcommand{\stagger}{\textsc{stagger}}
\newcommand{\balder}{\textsc{balder}}
\newcommand{\multitd}{\textsc{multi3d}}
\newcommand{\multi}{\textsc{multi}}
\newcommand{\marcs}{\textsc{marcs}}
\newcommand{\atmo}{\textsc{atmo}}
\newcommand{\breidablik}{\texttt{Breidablik}}

\begin{document}
\title{3D NLTE spectral line formation of lithium in late-type stars}
\author[E.~Wang, T.~Nordlander, M.~Asplund, A.~M.~Amarsi, K.~Lind, Y.~Zhou]{Ella~Xi~Wang$^{1, 2}$\thanks{Email: ellawang@mso.anu.edu.au},
Thomas~Nordlander$^{1, 2}$,
Martin~Asplund$^{3}$,
Anish~M.~Amarsi$^4$, \newauthor
Karin~Lind$^5$,
Yixiao~Zhou$^{1}$ \\
$^1$Research School of Astronomy and Astrophysics, Australian National University, Canberra, ACT 2611, Australia\\
$^2$ARC Centre of Excellence for All Sky Astrophysics in 3 Dimensions (ASTRO 3D), Australia\\
$^3$Max Planck Institute for Astrophysics, Karl-Schwarzschild-Str. 1, D-85741 Garching, Germany\\
$^4$Theoretical Astrophysics, Department of Physics and Astronomy, Uppsala University, Box 516, SE-751 20 Uppsala, Sweden \\
$^5$Department of Astronomy, Stockholm University, AlbaNova University Center, SE-106 91 Stockholm, Sweden
}
\maketitle

\begin{abstract}
Accurately known stellar lithium abundances may be used to shed light on a variety of astrophysical phenomena such as Big Bang nucleosynthesis, radial migration, ages of stars and stellar clusters, and planet engulfment events.
We present a grid of synthetic lithium spectra that are computed in non-local thermodynamic equilibrium (NLTE) across the \stagger\ grid of three-dimensional (3D) hydrodynamic stellar atmosphere models. This grid covers three Li lines at 610.4\,nm, 670.8\,nm, and 812.6\,nm for stellar parameters representative of FGK-type dwarfs and giants, spanning $\teff=4000$--7000\,K, $\logg=1.5$--5.0, $\met = -4.0$--0.5, and $\ali = -0.5$--4.0.
We find that our abundance corrections are up to 0.15\,dex more negative than in previous work, due to a previously overlooked NLTE effect of blocking of UV lithium lines by background opacities, which has important implications for a wide range of science cases.
We derive a new 3D NLTE solar abundance of \bb{$\ali = 0.96 \pm 0.05$, which is 0.09\,dex lower than the commonly used} value.
We make our grids of synthetic spectra and abundance corrections publicly available through the \breidablik\ package. This package includes methods for accurately interpolating our grid to arbitrary stellar parameters through methods based on Kriging (Gaussian process regression) for line profiles, and MLP (Multi-Layer Perceptrons, a class of fully connected feedforward neural networks) for NLTE corrections and 3D NLTE abundances from equivalent widths, achieving interpolation errors of the order 0.01\,dex.
\end{abstract}


\section{Introduction}
Lithium is the only element which can be produced through Big Bang nucleosynthesis, cosmic ray spallation, and stellar processes. Lithium is also a fragile element, destroyed in low-mass stars in a manner that depends on stellar mass, metallicity, age, and possibly other factors. The many production and destruction processes make lithium a complex but insightful element. Lithium can thus illuminate a number of distinct science cases, including the cosmological lithium problem, radial migration, ages of stars and stellar clusters, and planet engulfment events. However, these applications in large part rely on accurate measurements of lithium abundances in late-type stars.

The lithium abundance ($\ali$\footnote{We use the customary abundance notation where $\ax \equiv \log(N_\mathrm{X} / N_\textrm H) + 12$ and $[\mathrm{X}/\mathrm{Y}] \equiv (\textrm A(\mathrm{X})-\textrm A(\mathrm{Y})) - (\textrm A(\mathrm{X})-\textrm A(\mathrm{Y}))_\odot$, with $N_\mathrm{X}$ representing the number density of element ``X''.}) in stars can be used to estimate the cosmological value produced during Big Bang nucleosynthesis (\citealp[BBN;][]{cyburt16}).
Warm {old, metal-poor} dwarfs near the main-sequence turnoff-off (MSTO) have been found to exhibit roughly the same lithium abundances over a wide range of metallicities, the so-called Spite plateau \citep{spite82}. This plateau has $\ali \approx 2.0 - 2.2$, and is commonly interpreted to reflect the lithium abundance with which these old metal-poor stars were born. However, the cosmological lithium abundance predicted from BBN is $\ali = 2.75 \pm 0.02$, a factor of at least three larger than the abundance in the oldest stars \citep{pitrou18}.
The reason for this discrepancy is generally thought to be due to a combination of depletion and non-destructive deposition through gravitational settling \citep[e.g.][]{richard05} but speculation also exists that this may signal non-standard particle physics affecting the Big Bang nucleosynthesis production \citep{fields11}.
Parameterised stellar Li depletion predictions reproduce observations of lithium in globular cluster stars quite well, including the signature of dredge-up through a short-lived increase of lithium at the base of the subgiant branch \citep[e.g.][]{korn_atomic_2007,lind_atomic_2008,nordlander_atomic_2012,gruyters_atomic_2014,gruyters_atomic_2016}.
\citet{mucciarelli12} recognised that lithium abundances in metal-poor lower RGB stars also exhibit a plateau. These abundances are sensitive only to effects of dilution (reflecting the depth of the convection zone) and destruction that occurred prior, while effects of non-destructive deposition are erased by the deep convection zone.
Measurements for RGB stars in the Milky Way, $\ali \approx 1.0$ \citep{mucciarelli12}, as well as in the extragalactic globular cluster M54, $\ali \approx 0.9$ \citep{mucciarelli14}, imply an initial composition with $\ali \approx 2.3$--2.4 that does not vary with environment and therefore is unlikely to be depleted through external mechanisms. Although slightly higher than the Spite plateau, these estimates still imply significant destruction relative to the cosmological abundance.

At metallicities $\met > -1$, the abundance of lithium becomes coupled to $\met$ through chemical enrichment processes such as cosmic ray spallation \citep{prantzos12, prantzos17} and possibly also stellar production \citep{bensby18}.
Lithium abundance measurements in stars born at different locations and times can therefore tell us about the history of chemical evolution in the Galaxy, and radial migration of stars in the Galaxy \citep{schonrich09, prantzos17}.

Lithium abundance measurements can be used to determine ages for young stellar clusters.
Because pre-main sequence stars are fully convective, they will rapidly deplete lithium when their cores reach $T \approx 2.5 \times 10^6$\,K \citep{basri96, chabrier96, bildsten97, ushomirsky98, jeffries14}.
Even in coeval groups, the amount of depletion does however vary significantly from star to star \citep[e.g.][]{sestito05,zerjal19}, possibly tied to details of rotation and angular momentum transport \citep[see e.g.][]{donascimento09,soderblom10}.
It is therefore still unclear just why stars like the Sun exhibit significant lithium depletion by more than two orders of magnitude \citep[e.g.][]{melendez_solar_2010}. Similar patterns are seen in other stars, where it is still unclear whether this depletion correlates with the presence of exoplanets \citep{gonzalez08, israelian09, gonzalez15, delgado15} as several studies find no such correlation \citep{ryan00, baumann10, ramirez12, bensby18, carlos19}.
Some of the proposed mechanisms for enhanced lithium depletion for stars with planets are stellar angular momentum loss due to planetary migration \citep{castro08}, and strong differential rotation caused by interactions between the protoplanetary disk and star \citep{bouvier08}. Where no enhanced lithium depletion is detected, lithium depletion is instead linked to stellar ages; with older stars exhibiting increased lithium depletion \citep{baumann10, carlos19}.

In one-dimensional (1D) hydrostatic atmospheres, the accuracy of abundance determinations in late-type stars may be severely impacted by the commonly used approximations of local thermodynamic equilibrium (LTE). Modelling of lithium spectra in non-LTE (NLTE) has already been performed in many studies across a wide parameter space \citep[e.g.,][]{steenbock84,carlsson94,pavlenko96,takeda05,lind09,osorio11,takeda19}. These studies find that the The NLTE effect can lead to abundance corrections of up to 0.4\,dex \citep{lind09}.

In addition to LTE assumptions, 3D hydrodyamic atmospheres also affect measured lithium abundances. Recent studies suggest that although the differences between 3D NLTE and 1D LTE synthetic spectra have been found to often be rather small due to fortuitous cancellation between NLTE and 3D hydrodynamic effects that work in opposite directions, this cancellation is not perfect and abundance corrections may still be as large as 0.3\,dex \citep{klevas18}. Full 3D NLTE calculations  \citep[e.g.][]{kiselman97, asplund03, barklem03, sbordone10, lind13, klevas16, mott17, harutyunyan18, klevas18, mott20} have so far been limited in their coverage of parameter space due to the large computational cost involved.

This study presents a 3D NLTE Li grid spanning the full parameter range expected for FGK-type dwarfs and giants, covering a subset of $\teff=4000$--7000\,K, $\logg=1.5$--5.0, $\met = -4$--0.5 -- a total of 195 3D hydrodynamic model atmospheres -- with abundances in the range $\ali=-0.5$--4. We detail the model atom and stellar atmospheres in Section~\ref{sec:code}. In Section~\ref{sec:res}, we present and discuss our results with comparison to previous work. We introduce our interpolation methods and the publicly available synthetic spectra, spectrum interpolation, and abundance correction package \breidablik\footnote{\url{https://github.com/ellawang44/Breidablik}} in Section~\ref{sec:int}.
Lastly, in Section~\ref{sec:dis} we present results of our new lower inferred abundances on a handful of science cases, before presenting our conclusions in Section~\ref{sec:con}.

\section{3D NLTE Spectral Line Formation for Li}
\label{sec:code}
We use the 3D NLTE radiative transfer code \balder, which originates in the \multitd\ code \citep{botnen99,leenaarts09} but has since been developed significantly, including a new equation-of-state and opacity package \blue\ \citep{amarsi16a,amarsi16b,amarsi18}.

The statistical equilibrium is solved by calculating the mean radiation field with short characteristic rays. These are distributed as $\mu \equiv \cos \theta$ relative to the vertical on the interval $-1$ to $+1$ using a Gauss-Lobatto quadrature with 8 points.
For non-vertical rays, we consider four equidistant azimuthal angles, giving a total of 26 rays.
The emergent flux is computed by calculating the emergent intensity in the vertical direction, as well as in seven inclined directions each using eight horizontal angles, for a total of 57 rays.
We compute spectra for abundances in the range $\ali = -0.5$ to $+4.0$, in steps of 0.5\,dex, \bb{which produces an abundance interpolation error of at most $\sim 0.02$\,dex, as determined through comparison to calculations over the full abundance range with very small steps.}

\subsection{Li Model Atom}
\label{sec:atom}
\begin{figure}
	\centering
	\includegraphics[width=0.5\textwidth]{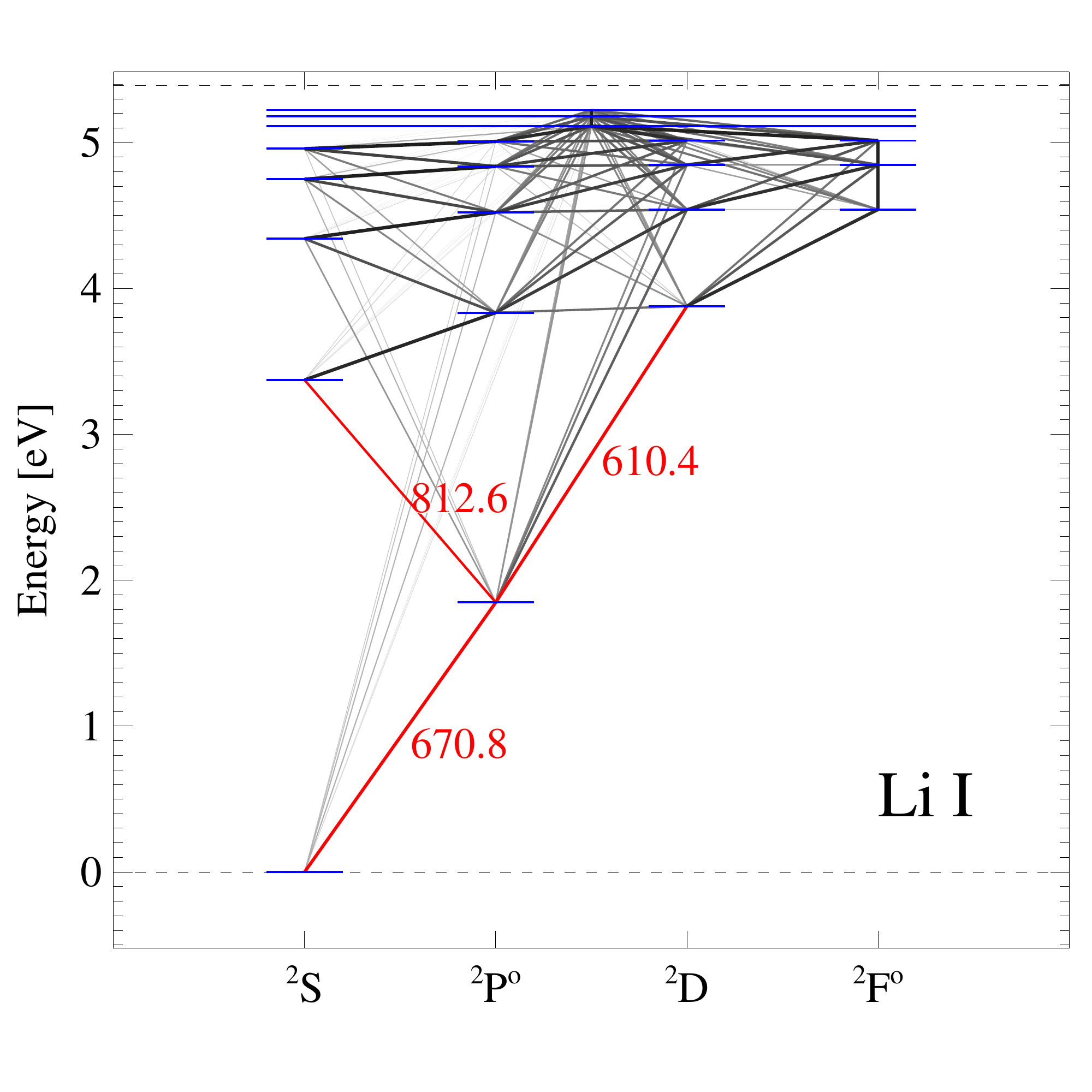}
	\caption{Term diagram of lithium, illustrating the energies of different states, ordered according to spectroscopic term. Bound-bound radiative transitions are marked with black lines, and the abundance diagnostic lines at 610.4\,nm, 670.8\,nm, and 812.6\,nm are highlighted in red. The long horizontal blue lines are super levels where the $l$ quantum number has been collapsed. The $^2\textrm{F}^0$ states are superstates representing $l \ge 3$.}
	\label{fig:termdiag}
\end{figure}

We use a model atom containing 20 levels of \ion{Li}{i} plus the \ion{Li}{ii} ground state, connected by 113 bound-bound and 20 bound-free transitions. We compute the spectra of the three strongest transitions, highlighted in Fig.~\ref{fig:termdiag}.

The model atom originates with \citet{carlsson94}, and was substantially updated by \citet{lind09} and \citet{osorio11}. Briefly, energy levels and radiative transition data come from TOPbase \citep{peach88}, with some notable exceptions.
The atomic data for the 670.8\,nm transition uses an oscillator strength from \citet{yan98}, fine structure splitting from \citet{sansonetti11} and hyperfine splitting from \citet{beckmann74} and \citet{puchalski09}. Although updated, the resulting wavelengths and \bb{relative oscillator strengths} are consistent with those of \citet{smith98}.
Likewise, the 610.4\,nm transition uses an oscillator strength from \citet{yan98} and fine structure splitting from \citet{lindgard77}. For all transitions, we assume the presence of $^7$Li only, neglecting any isotopic splitting.
The three strongest transitions use collisional broadening parameters based on \citet{anstee95} and \citet{barklem97}.

Cross-sections for excitation and ionisation through electron collisions were calculated and implemented by \citet{osorio11}.
Inelastic hydrogen collisional transition rates for excitation and charge transfer of low-lying states \citep{barklem03} are based on cross-sections from \citet{croft_rate_1999} and \citet{belyaev03}.
We have further implemented inelastic hydrogen collisional excitation rates for more highly excited states following \citet{kaulakys85,kaulakys91}, using the publicly available code KAULAKYS \citep{barklem16}.

\subsection{Stellar atmospheres}
\subsubsection{3D Hydrodynamical Stellar Model Atmospheres}
\begin{figure}
	\centering
	\includegraphics[width=0.5\textwidth]{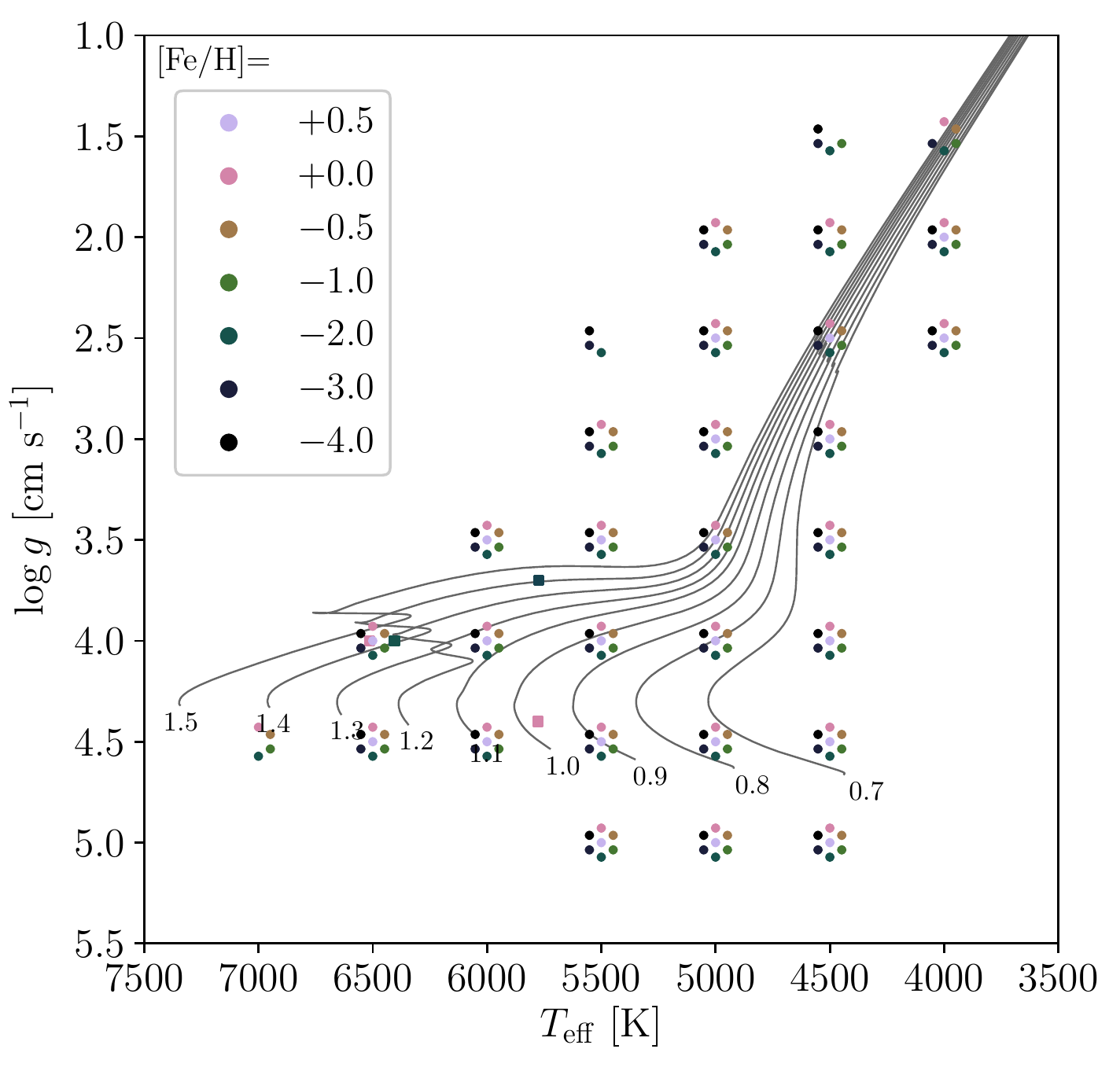}
	\caption{Stellar parameters of the Stagger-grid models \citep{magic13} used in this work, colour-coded according to $\met$;
	the Sun, HD 84937, HD 140283 and Procyon are shown with squares.
	Stellar evolution tracks at solar metallicity with masses in the range 0.7--1.5\,$\msun$ in steps of 0.1\,$\msun$ from MIST \citep{paxton11, paxton13, paxton15, choi16, dotter16} are shown for reference, labeled according to mass.}
	\label{fig:stagger}
\end{figure}

To calculate 3D NLTE synthetic spectra, we use \balder\ to perform radiative transfer post-processing of 3D hydrodynamical model atmospheres from the \stagger-grid \citep{magic13}.
Each simulation covers a time sequence of roughly two convective turnover times, represented by about 150 snapshots. A small number of simulations suffered from convergence problems, and are therefore not considered in this work.
As a result, we selected 195 models, shown in Fig.~\ref{fig:stagger}. These cover stellar parameters representing FGK-type dwarfs and giants, in a wide range of metallicities between $\met = -4.0$ and $+0.5$. All models adopt the \citet{asplund09} metal mixture, aside from models with $\met \le -1$ where an $\alpha$-enhancement of $[\alpha/\rm{Fe}] = 0.4$ dex is further applied.

From each simulation, we select five temporally equidistant snapshots. Test calculations based on larger sets of snapshots indicate that using five snapshots results in a typical abundance error ($\sigma / \sqrt N$) less than 0.02\,dex and 0.01\,dex in LTE and NLTE respectively based on temporal variations in the equivalent widths. More specifically, we find the largest snapshot-to-snapshot variations amongst the hottest models, with equivalent widths varying by typically $\sigma = 0.03$\,dex in LTE and by $0.02$\,dex in NLTE. Importantly, the ratio of LTE and NLTE equivalent widths may have a larger standard deviation, of the order 0.04\,dex, indicating that temporal variations may have different sign in LTE and NLTE and do not perfectly correlate. This thus indicates that fewer snapshots are required to sample a simulation with NLTE calculations than the corresponding LTE calculations.
These test calculations furthermore suggest that one should avoid the approach where NLTE corrections are derived from a small number of snapshots and applied to a longer sequence of LTE calculations, aside from possible issues related to differences in line shapes.

The original 3D hydrodynamic simulations were computed on a staggered Cartesian grid with $240^3$ elements, extending deep below the stellar photosphere. However, for the radiative post-processing, we interpolate the hydrodynamic simulations to $80 \times 80$ volume elements in the horizontal direction and 220 volume elements along the vertical and truncate the deep layers. This interpolation retains refined resolution of the continuum-forming regions that have steep temperature gradients, as well as the optically-thin line-forming regions. The truncation and interpolation follows that described in Section~2.1 of \citet{amarsi18}.
The total number of volume elements is therefore of the order $10^6$.

\subsubsection{1D Hydrostatic Stellar Model Atmospheres}
For comparison, we also use a set of custom 1D hydrostatic stellar model atmospheres which treat convection using the classical mixing length theory computed with the \atmo\ code written by W. Hayek (see Appendix A of \citet{magic13} for details).
The 1D models were computed at exactly the same stellar parameters as the 3D hydrodynamic models, share the same equation of state and opacities, and were based on the same opacity binning procedure.
We also compute additional \bb{1D} models at randomly selected values of $\teff$ and $\logg$ in order to test our interpolation procedures.

Unlike the 3D hydrodynamic models, where convective motions arise naturally from first principles, convection in these 1D models is approximated using mixing length theory controlled by a mixing length parameter ($\mlt \equiv l / H_{\mathrm P}$, where $l$ is the mixing length and $H_\mathrm{P}$ is the pressure scale height). We used models computed with $\mlt = 1.0$, 1.5, and 2.0, thereby covering a range typically used in the literature. In the radiative transfer post-processing, small-scale hydrodynamic motions that produce an effective broadening of line opacities leading to desaturation of spectral lines are mimicked through a local broadening parameter known as microturbulence \citep{asplund00}. We select the values $\vmic = 0$, 1, and 2\,\kms. For our default comparisons, we use models with $\mlt = 1.5$ and $\vmic = 1$\,\kms.

In addition to these custom model atmospheres, we also use the commonly used \marcs\ grid \citep{gustafsson08}, which covers a wide range in stellar parameters representative of FGKM-type dwarfs and giants.

\section{Results}
\label{sec:res}
In this section, we introduce the main NLTE effects on Li as discussed in the literature based on 1D (Section~\ref{sec:res1D}) and 3D (Section~\ref{sec:res3D}) modelling.
We present our 3D line formation results in Section~\ref{sec:our_work}. In Section~\ref{sec:UV}, we compare our results to previous work, and discuss a previously neglected effect of background line opacity blocking of lithium lines in the UV that would otherwise deplete low-lying populations through photon pumping. With this effect taken into account, we generally find stronger optical lines implying lower inferred lithium abundances. Lastly, in Section~\ref{sec:nltec}, we discuss the abundance correction over the HR diagram.

\subsection{Departures from LTE in 1D models}
\label{sec:res1D}
The main NLTE mechanisms relevant to lithium have been presented in a large body of previous studies
\citep[e.g.,][]{steenbock84,carlsson94,pavlenko96,takeda05,lind09,osorio11}.
While our model represents a substantial upgrade in terms of, e.g., collisional transition rates and the treatment of background opacities, the pedagogical description of NLTE mechanisms and behaviours by \citet{carlsson94} largely still applies to our results.
As discussed by \citet{lind09}, this agreement results from newer model atmospheres exhibiting steeper temperature gradients that increase the excess of ultraviolet radiation and thus leads to enhanced over-ionisation, which partially cancels with the use of hydrogen collisional transition rates from \bb{\citet{barklem03} that are significantly larger than the previously used hydrogen collisional rates predicted through the \citet{drawin_zur_1968} formula.}
Furthermore, while \citet{osorio11} found large differences between their R-matrix calculations of electron collisional transition rates compared to previous semi-empirical estimates, the two happen to show agreement within a factor of two for the dominant 2s--2p resonance transition, resulting in negligible effects on level populations and abundance corrections.

When the the 670.8\,nm resonance line is strong, departures from LTE are mainly controlled by the line itself through photon losses in so-called resonance scattering. \bb{In NLTE, these photon losses dictate that the radiation field ($J_\nu$) in the line becomes sub-thermal}, i.e. smaller than the Planck function ($B_\nu$).
For a sufficiently strong line where line opacity dominates over continuum opacity, the source function ($S_\nu$) becomes controlled by the line rather than the continuum. As the line source function ($S_l$) becomes tied to the radiation field ($S_l \approx J_\nu < B_\nu$), the drop in $S_l$ causes core darkening that strengthens the line, thus leading to lower inferred abundances in NLTE compared to LTE.

The steep temperature dependence of $B_\nu$ in the ultraviolet may produce a super-thermal radiation field ($J_\nu > B_\nu$), which has the opposite effect to resonance scattering. Photoionisation through the continua of the ground and first excited states (2s and 2p) causes a drop in the populations of \ion{Li}i through over-ionisation in the presence of steep temperature gradients or small ultraviolet opacities. The corresponding effect of photon pumping through ultraviolet spectral lines rather than continua is largely quenched by background metal line opacities.

At infrared wavelengths,
\bb{the opposite case will typically be true, with a sub-thermal radiation field
($J_\nu < B_\nu$)}. The resulting deficit of photoionisation of highly excited states leads to increased level populations through over-recombination.
The flow of excess electrons propagates down through photon losses in a chain of infrared lines, and further to the ground state through the resonance line.

The interplay of these various NLTE mechanisms leads to abundance corrections that vary in magnitude and sign with stellar parameters as well as the abundance of lithium.
A stronger resonance line will produce more negative abundance corrections through resonance scattering. Higher $\teff$, and to lesser extent lower $\logg$, weakens the resonance line and further makes abundance corrections more positive through enhanced overionisation.
At lower $\met$ the smaller background opacities compete with a flattening of the atmospheric temperature gradients due to a lack of radiative cooling (in 1D model atmospheres), that in practice leads to only a small enhancement of overionisation and thus abundance corrections that become slightly more positive.

\bb{The atom presented in Section~\ref{sec:atom} was also used to compute a comprehensive grid of 1D departure coefficients for use in GALAH DR3 (Buder et al., in prep), and made available through \citep{anish20, anish_mayur_amarsi_2020_3982506}.}

\subsection{Departures from LTE in 3D models}
\label{sec:res3D}
While the mechanisms outlined above are still the dominant ones in 3D model atmospheres, their relative importance and behaviour with stellar parameters changes due to the generally steeper temperature gradients present.

In particular, the outer layers of hydrodynamic atmospheres are effectively cooled by convective motions and adiabatic expansion. In warm metal-poor models, the deficit of opacities leads to radiative heating and cooling rates far below equilibrium, which alongside the convective cooling mechanism, results in significantly lower temperatures than seen in hydrostatic atmospheres in radiative equilibrium \citep{asplund99,collet06}.
In LTE, low temperatures in the outer regions of the atmospheres lead to a decrease in the ionisation fraction of lithium and thus strongly enhanced populations and resulting line strengths, in particular in the steep temperature structures formed by hot upwelling gas in granules, producing negative abundance corrections of the order 0.3\,dex in warm metal-poor dwarfs \citep{asplund99,klevas16}.

Pioneering work utilising limited radiative transfer in 3D hydrodynamic simulations of the Sun indicated that a decoupling of the radiation field from the local gas properties would drive significantly stronger overionisation effects than in the corresponding 1D hydrostatic models, weakening the line by 0.1\,dex \citep{kiselman97,uitenbroek98}. This decoupling was also observationally verified in resolved spectra of the solar surface \citep{kiselman98}, which in contrast to 3D LTE predictions exhibited no significant variation of line strengths as a function of the continuum intensity.
Calculations by \citet{asplund03} and \citet{barklem03} with the \multitd\ code \citep{botnen99} verified these early results, and extended work to metal-poor dwarfs where effects were significantly stronger with abundance corrections comparing 3D NLTE to 3D LTE of the order 0.2\,dex, resulting in good agreement between 3D NLTE and 1D NLTE calculations. They found that the effect of enhanced overionisation was particularly pronounced in regions where the temperature gradient is strong, i.e. where LTE line strengths were the most strongly enhanced, while the opposite behaviour may appear above intergranular lanes.

Further work has extended the coverage of 3D NLTE calculations to better sample the spatial and temporal variations in the 3D hydrodynamic simulations of metal-poor dwarfs \citep{sbordone10,lind13} and giants \citep{klevas16,nordlander17} as well as solar-metallicity stars \citep[e.g.][]{mott17,harutyunyan18}.
We note however that these calculations are heterogeneous and the results are sometimes incompatible due to different implementations of model atoms (in terms of electron structure complexity and collisional rates), radiative transfer and opacity treatment, and model atmospheres.

\subsection{3D NLTE line formation}
\label{sec:our_work}

\begin{figure}
	\centering
	\includegraphics[width=0.49\textwidth]{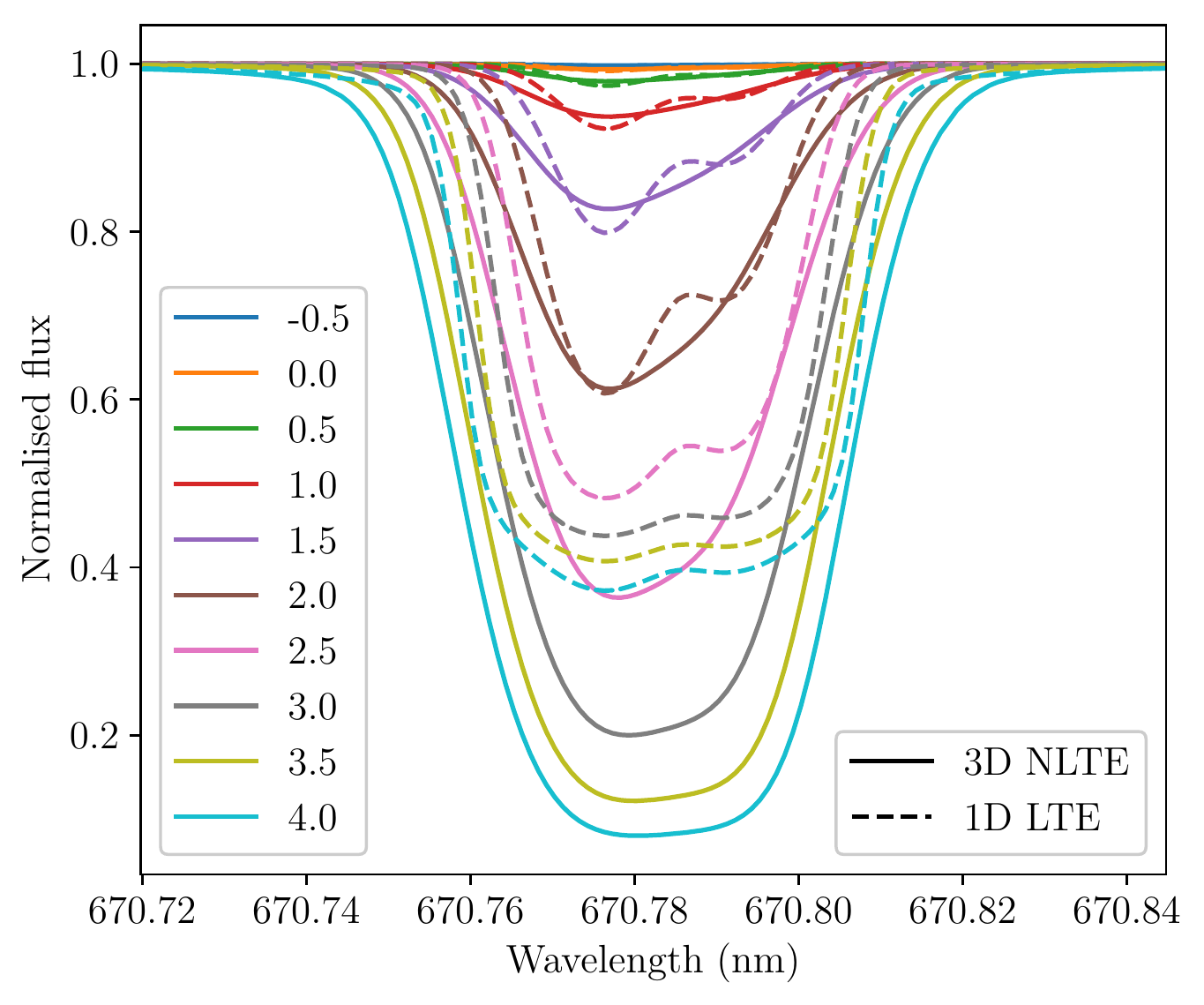}
	\caption{Line profiles computed in 3D NLTE (solid lines) and 1D LTE (dashed lines) for a cool red giant star model with nominal parameters $T_{\rm{eff}} = 5000$\,K, $\log g = 2.0$, $[\rm{Fe}/\rm{H}] = -2.0$, and a range of $\ali$ (colour-coded as indicated in the legend). The 1D model uses $\mlt = 1.5$ and $\vmic=1$\,\kms. Neither rotational nor macro-turbulent broadening has been applied.}
	\label{fig:line_profiles}
\end{figure}

We show in Fig.~\ref{fig:line_profiles} profiles of the 670.8\,nm resonance line for a very metal-poor model representing a star on the red giant branch.
While the 1D LTE line profiles are sharp with clear fine structure components, velocity fields in the 3D model of the order 5\,\kms\ (as compared to the thermal linewidth of 3\,\kms) cause a smoothed appearance for the 3D NLTE profiles.
At the lowest $\ali$, the line forms in relatively deep atmospheric layers
where high pressures lead to high collisional rates that nearly uphold LTE.
As $\ali$ increases, line formation begins to push higher in the photosphere toward layers with significantly lower temperatures in the hydrodynamic simulation than in the corresponding hydrostatic model.
More importantly however, photon losses in the core of the 670.8\,nm line become significant, leading to a deeper line with a NLTE effect that strengthens with increasing $\ali$ through so-called resonance scattering.

\begin{figure*}
	\centering
	\includegraphics[width=\textwidth]{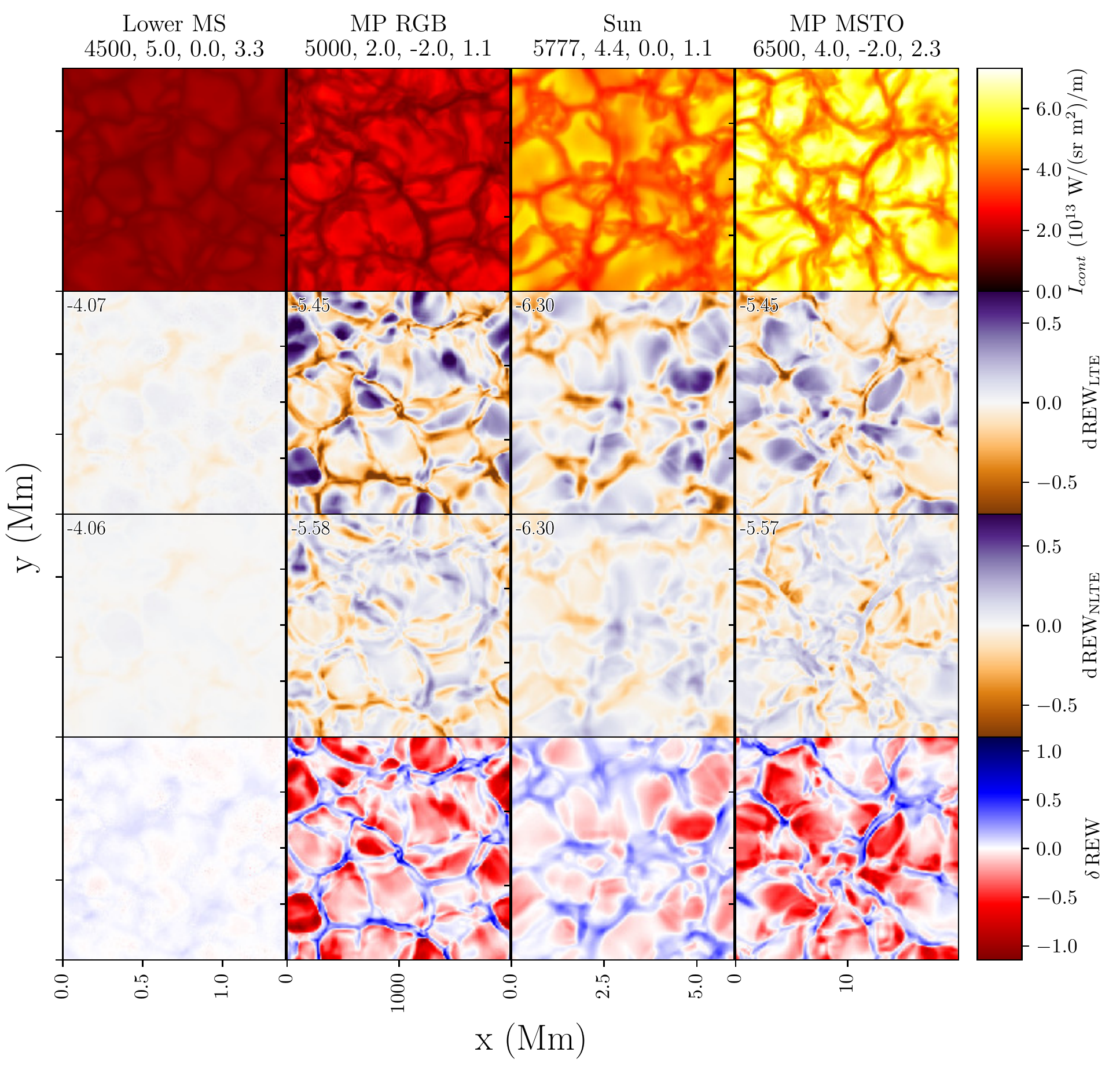}
	\caption{Top-down view of different 3D stellar models (columns), illustrating different properties of the 670.8\,nm transition in disk-centre intensity (rows). Models are labeled by their stellar parameters, $\teff$, $\logg$, $\met$ and $\ali$.
	From top to bottom, the rows illustrate the continuum intensity across the stellar surface at 670.8\,nm ($I_{\mathrm {cont}}$), the variation in LTE reduced equivalent width ($\rewlte$) relative to its average (d\,REW$_{\rm{LTE}}$), likewise in NLTE (d\,REW$_{\rm{NLTE}}$), and the difference between $\rewnlte$ and $\rewlte$ ($\delta \rm{REW}$).
	The average values of $\rewlte$ and $\rewnlte$ are labelled in the top left of the corresponding panels.
	}
    \label{fig:surf}
\end{figure*}

Fig.~\ref{fig:surf} shows top-down views of four representative model atmospheres, as seen in disk-centre intensity. From left to right, these represent a star on the lower main-sequence, a metal-poor red giant branch star, the Sun, and a metal-poor main-sequence turn-off star as commonly observed on the Spite plateau. Each row illustrates different properties of the 670.8\,nm line: the local continuum intensity ($I_{\mathrm cont}$), the variation of log reduced equivalent width (REW) relative to its average in LTE (d\,REW$_{\rm{LTE}}$) and NLTE (d\,REW$_{\rm{NLTE}}$), and finally the NLTE effect on line strengths. In this study, we use the reduced equivalent width:
\begin{equation}
\textrm{REW} = \log_{10} \left( \frac{\textrm{EW}}{\lambda_0} \right)
\label{eq:REW}
\end{equation}
where EW is the equivalent width of the line profile and $\lambda_0$ its central wavelength at rest.
The last row of Fig.~\ref{fig:surf} shows the difference in line strength between NLTE and LTE, defined as
\begin{equation}
\delta \rm{REW} = \rewnlte - \rewlte.
\end{equation}
While the dominant difference in continuum intensity is due to the variation in surface temperature and therefore overall brightness, differences in contrast can also be appreciated to increase with $\teff$ and decrease with $\logg$ and $\met$ \citep[for an in-depth discussion, see][]{magic13}.
Like previous work, we find that the LTE line strength anti-correlates with the surface intensity pattern, such that bright upflowing granules due to their strong temperature gradients typically exhibit strong lines, compared to the cool intergranular lanes with weaker temperature gradients.
In NLTE however, this correlation is weakened due to the influence of non-local radiation fields. Foremost, the bright granules illuminate the line-forming layers immediately above, leading to a strongly super-thermal radiation field. This drives an excess of photoionisation in the continuum of the first excited state, 2p, leading to over-ionisation that weakens the line strength \bb{relative to LTE} by as much as a factor of ten for the two metal-poor models. In contrast, the line forms deeper above intergranular lanes, where the flatter temperature gradient does not produce a strongly super-thermal radiation field, and instead sub-thermal radiation across the infrared continua of more highly excited states drive over-recombination.
A similar but milder effect of this is also seen in the solar model, where its weak line strength overall shifts line formation deeper. At the same time, its high surface gravity and strong metal line opacity both lessen all departures from LTE.

\begin{figure*}
	\centering
	\includegraphics[width=1.12\textwidth]{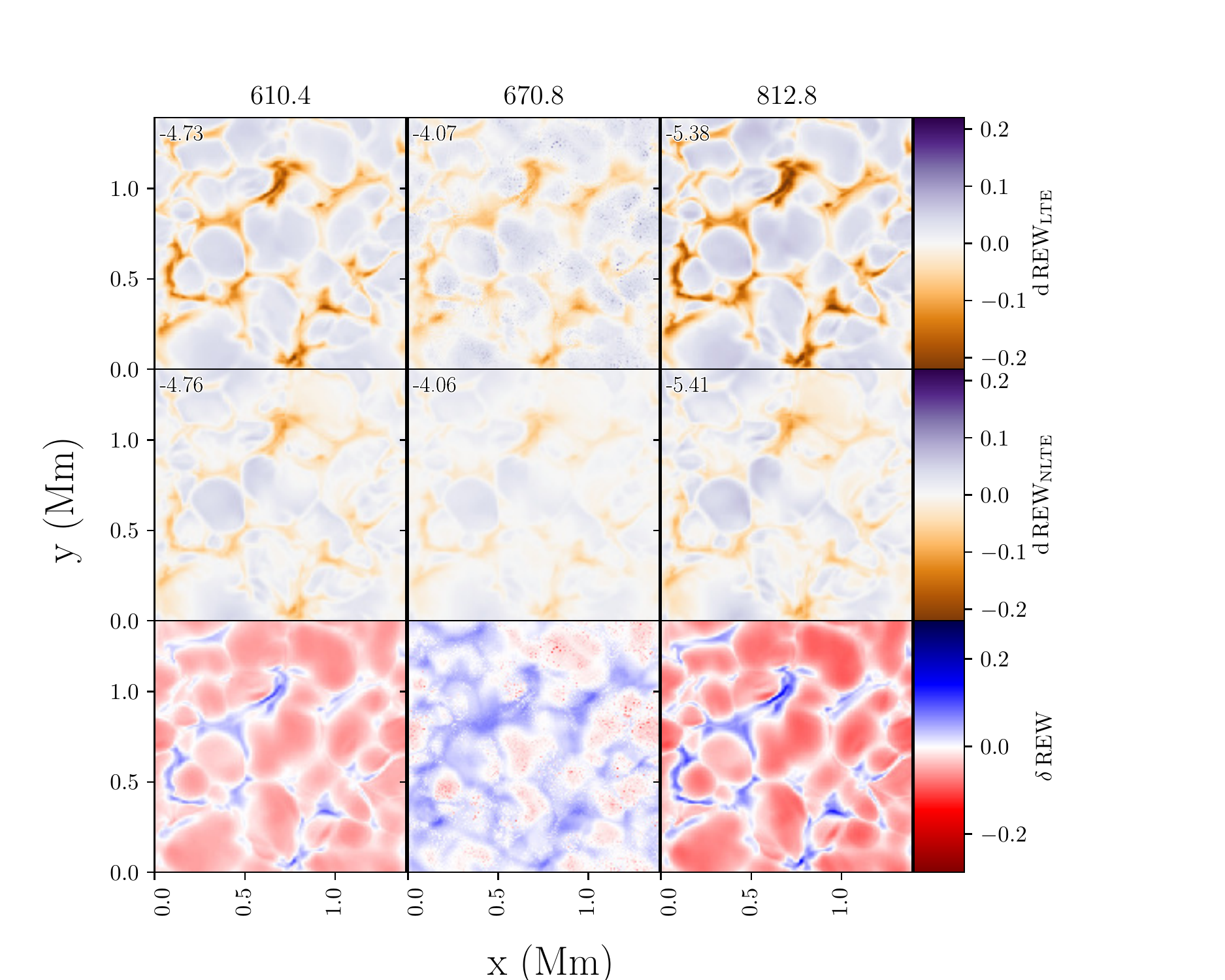}
	\caption{Top-down view of a 3D stellar model with $\teff = 4500$\,K, $\logg = 5.0$ and $\met = 0.0$, calculated with $\ali = 3.3$ for different spectral lines (columns, labeled by wavelength in units of nm). From top to bottom, the rows illustrate the variation in LTE reduced equivalent width ($\rewlte$) relative to its average (d\,REW$_{\rm{LTE}}$), likewise in NLTE (d\,REW$_{\rm{NLTE}}$), and the difference between $\rewnlte$ and $\rewlte$ ($\delta$\,REW). The average values of $\rewlte$ and $\rewnlte$ are labelled in the top left of the corresponding panels.}
	\label{fig:t45g50m00_surf}
\end{figure*}

The discussion above focused mainly on the three warm models. The cool lower main sequence model exhibits similar effects to the others, albeit at smaller magnitude due to the immense line strength of nearly 600\,m\AA, because it's equivalent width is proportional to $\sqrt{n}$ ($n$ is the population of lower state) instead of proportional to $n$ like for weak lines. The combination of low surface temperature and high $\ali$ however means the subordinate lines at 610.4 and 812.6\,nm become measurable.
We therefore look more closely at this model by examining its surface variation for all three lithium lines in Fig.~\ref{fig:t45g50m00_surf}.
The subordinate lines originate from the first excited state at $E_{\rm low} = 1.84$\,eV and are therefore significantly weaker than the resonance line, and thus probe different heights of formation as compared to the resonance line.
For the 670.8\,nm line, the core depth is mainly controlled by resonance scattering, which compared to LTE always enhances the line strength through core darkening. This results in the line core having similar depth across the entire surface in NLTE.
In LTE however, the core may be significantly weakened either due to the weak temperature gradient seen above intergranular lanes, or due to temperature \bb{contrast} inversions in the optically thin layers caused by mechanical heating.
In NLTE, the wings of the 670.8\,nm line as well as the cores of the 610.4\,nm and 812.6\,nm lines, are instead controlled by the excess of UV flux. There is a balance between over-ionisation and over-recombination that is driven in a similar way to that seen for the other models in Fig.~\ref{fig:surf}, and where on average over-ionisation dominates.

\subsection{The role of UV lines}
\label{sec:UV}

\begin{figure}
	\includegraphics[width=0.49\textwidth]{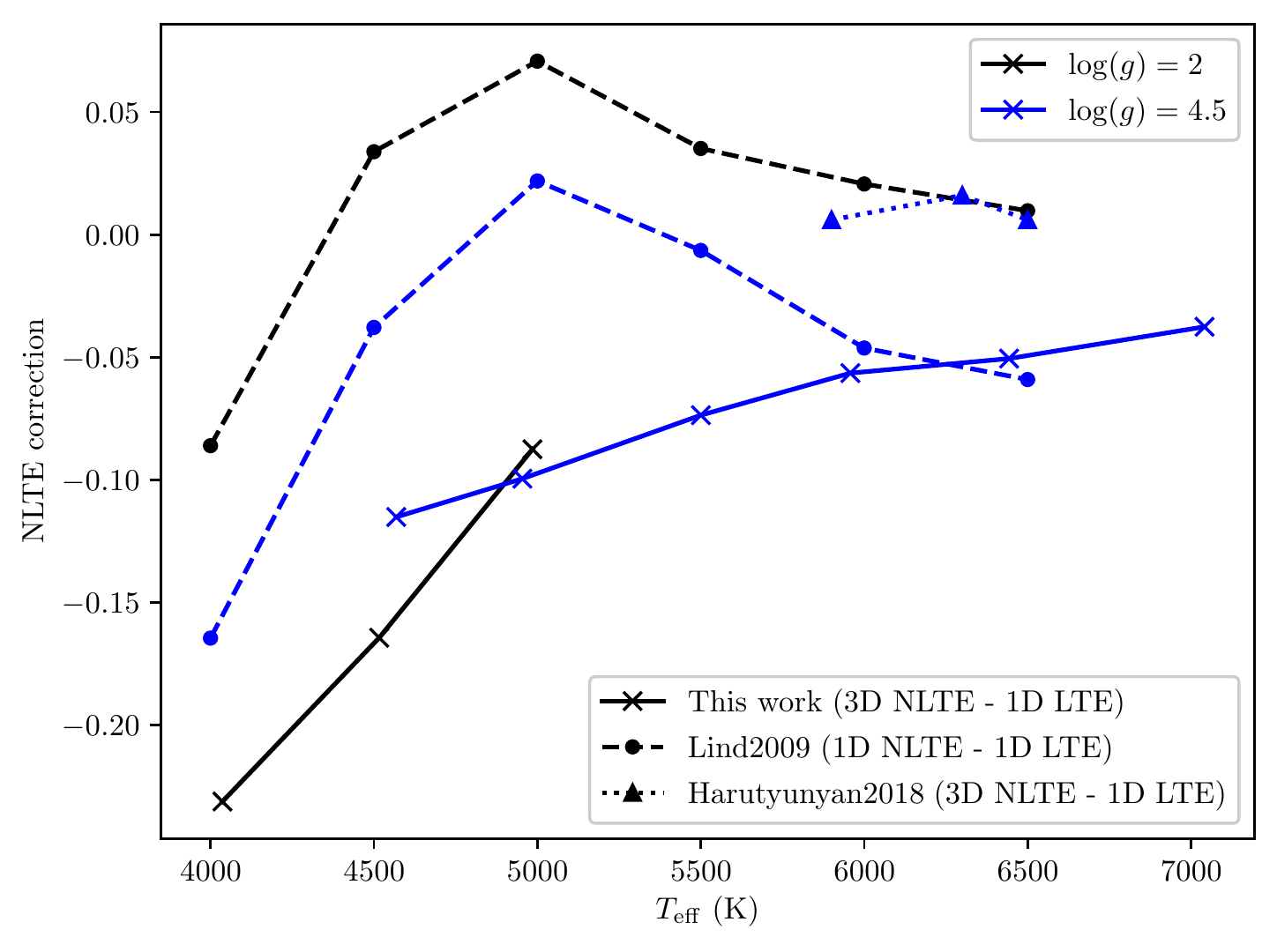}
\caption{NLTE correction for models with varying $\teff$, comparing this work (3D NLTE--1D LTE, solid line with crosses), \citet{lind09} (1D NLTE--1D LTE, dashed line with circles), and \citet{harutyunyan18} (3D NLTE--1D LTE, dotted lines with triangles). All models use $\met = -1$ and $\ali = 2$. We show models with $\logg = 2.0$ (in black) and $\logg = 4.5$ (in blue).
}
\label{fig:comp}
\end{figure}

In order to quantify NLTE effects, we define abundance corrections,
\begin{equation}
\Dn = \textrm{A(Li)}_{\textrm{3D,NLTE}} - \textrm{A(Li)}_{\textrm{1D,LTE}},
\end{equation}
where $\ali$ is calculated based on matching REW.
Our $\Dn$ are generally more negative than found in previous work, as shown in Fig.~\ref{fig:comp}, with differences being larger at lower $\teff$ and $\logg$, and higher $\met$.
This is an effect of our more accurate treatment of background line opacities: In \balder, overlapping transitions are implemented through the use of a common set of wavelengths and opacities for all line and continuum transitions, including the background lines and continua from other elements that are treated in LTE. After computing the statistical equilibrium, background line opacities can optionally be switched on or off in order to produce emergent spectra with or without blends, while overlapping transitions in the atom of study are always retained unless individual transitions are explicitly disabled.
This is in contrast to, e.g., the \multi\ code \citep{carlsson86}, which does not treat overlapping transitions and by default only considers background line opacities that blend with continua but not line transitions \citep[see e.g.][]{collet05}, unless blending line transitions are specified explicitly \citep[e.g. as was done by][]{nordlander17al}.

\begin{figure}
	\includegraphics[width=0.49\textwidth]{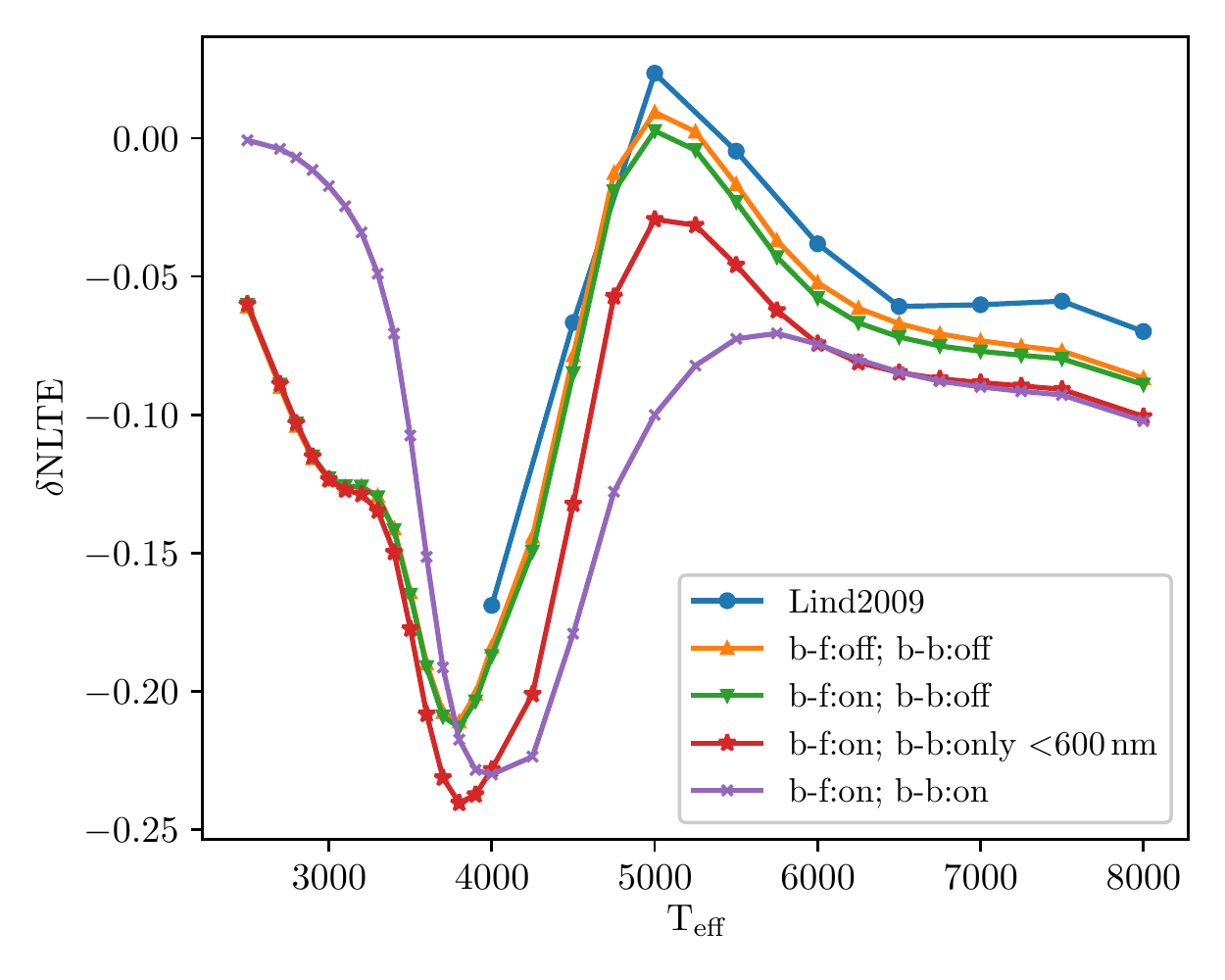}
    \caption{Test calculations illustrating the 1D NLTE abundance correction ($\dn$) for MARCS models with $\logg = 4.0$, $\met = -1.0$, and $\ali = 2.0$. Models have been computed using the model atom from \citet{lind09} with background line-opacities enabled or disabled for bound-free (b-f) and line (b-b) transitions of lithium (colours according to figure legend); results from \citet{lind09} are shown for comparison (in blue).}
\label{fig:back_op}
\end{figure}

We illustrate effects of partially disabling background line opacities in Fig.~\ref{fig:back_op}, where we compare results from \citet{lind09} to test calculations with MARCS models using the same model atom as in their work.
As expected, we find that removing background line blocking for lines redward of 600\,nm mainly causes a decrease of photon losses in the 670.8\,nm line, which is itself normally responsible for increasing the ground state population through so-called resonance scattering.
When disabling background line blocking of all line transitions, we find that a number of UV resonance and subordinate lines together deplete the 2s and 2p states through photon pumping by an amount comparable to that of UV photoionisation, resulting in a significant weakening of all low-excitation lines.
In contrast, through our consistent inclusion of background line opacities, we find that radiative excitations through UV line transitions are effectively quenched, having radiative brackets (the net difference between in-going and out-going radiative transition rates) that are effectively zero.
We note that \citet{carlsson94} found that photon pumping through UV lines was not important -- most likely this is due to their use of older model atmospheres with generally flatter temperature structures that produced less superthermal radiation.

\subsection{3D NLTE abundance corrections}
\label{sec:nltec}

\begin{figure*}
	\centering
	\includegraphics[width=0.4\textwidth]{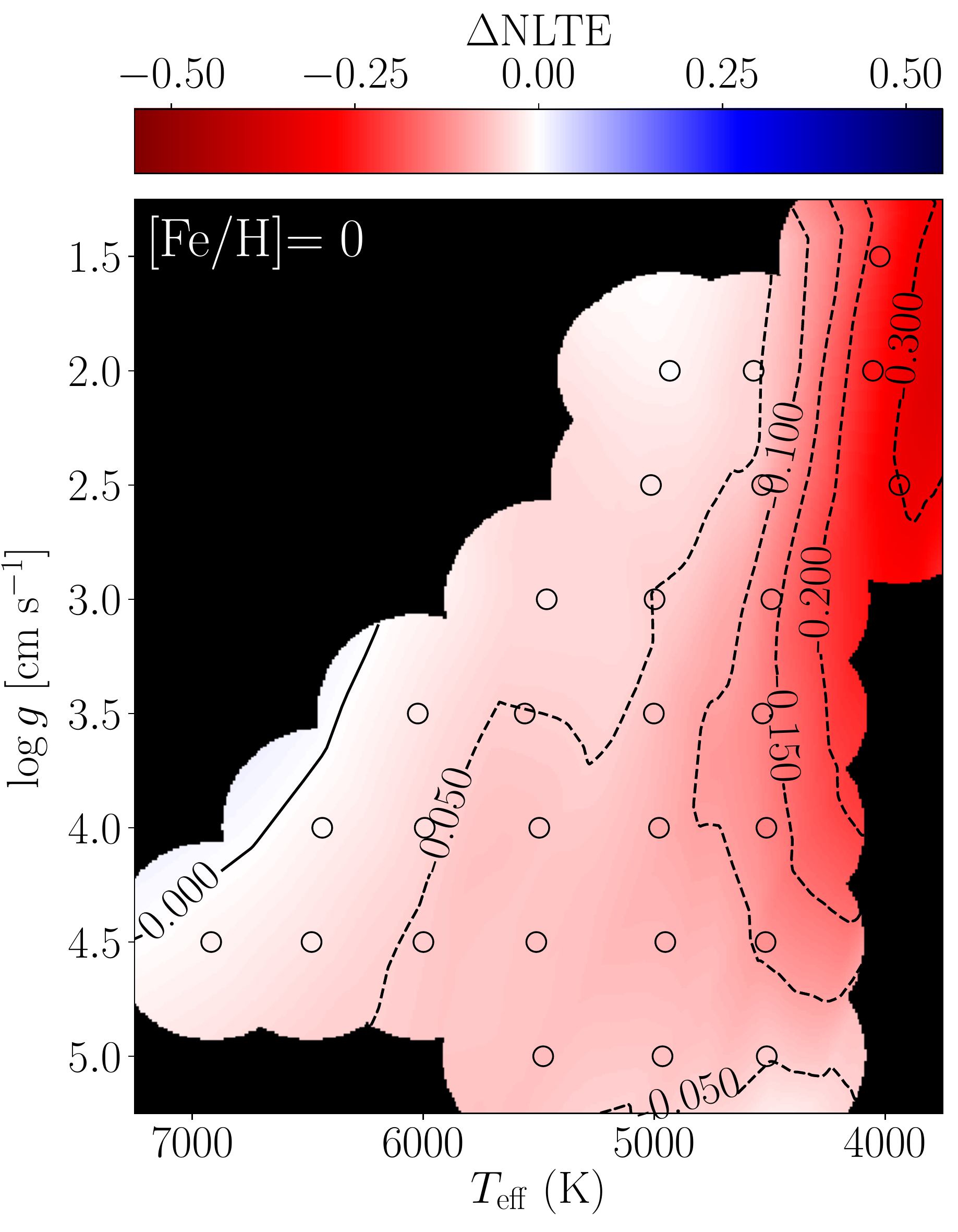}
	\includegraphics[width=0.4\textwidth]{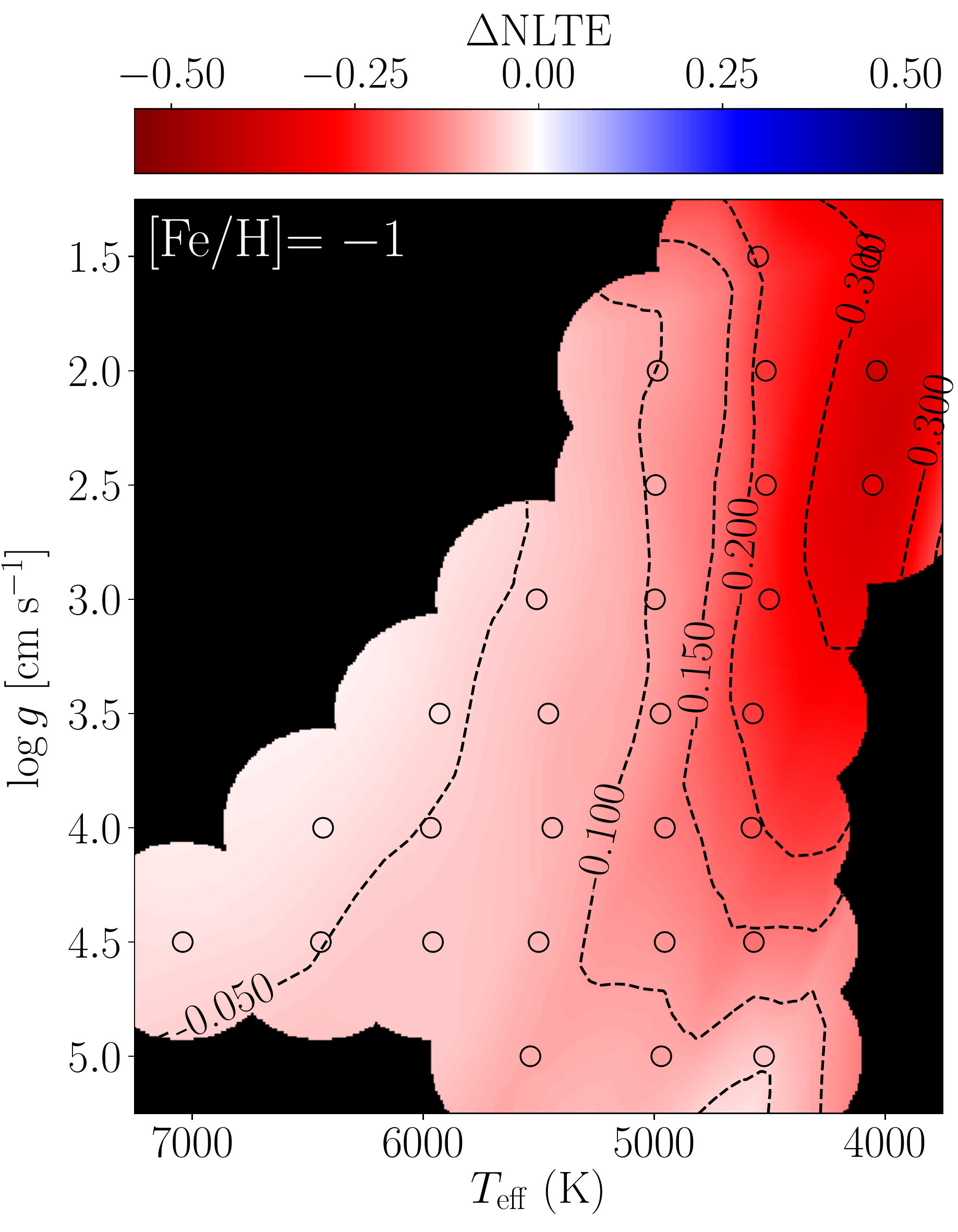}
	\includegraphics[width=0.4\textwidth]{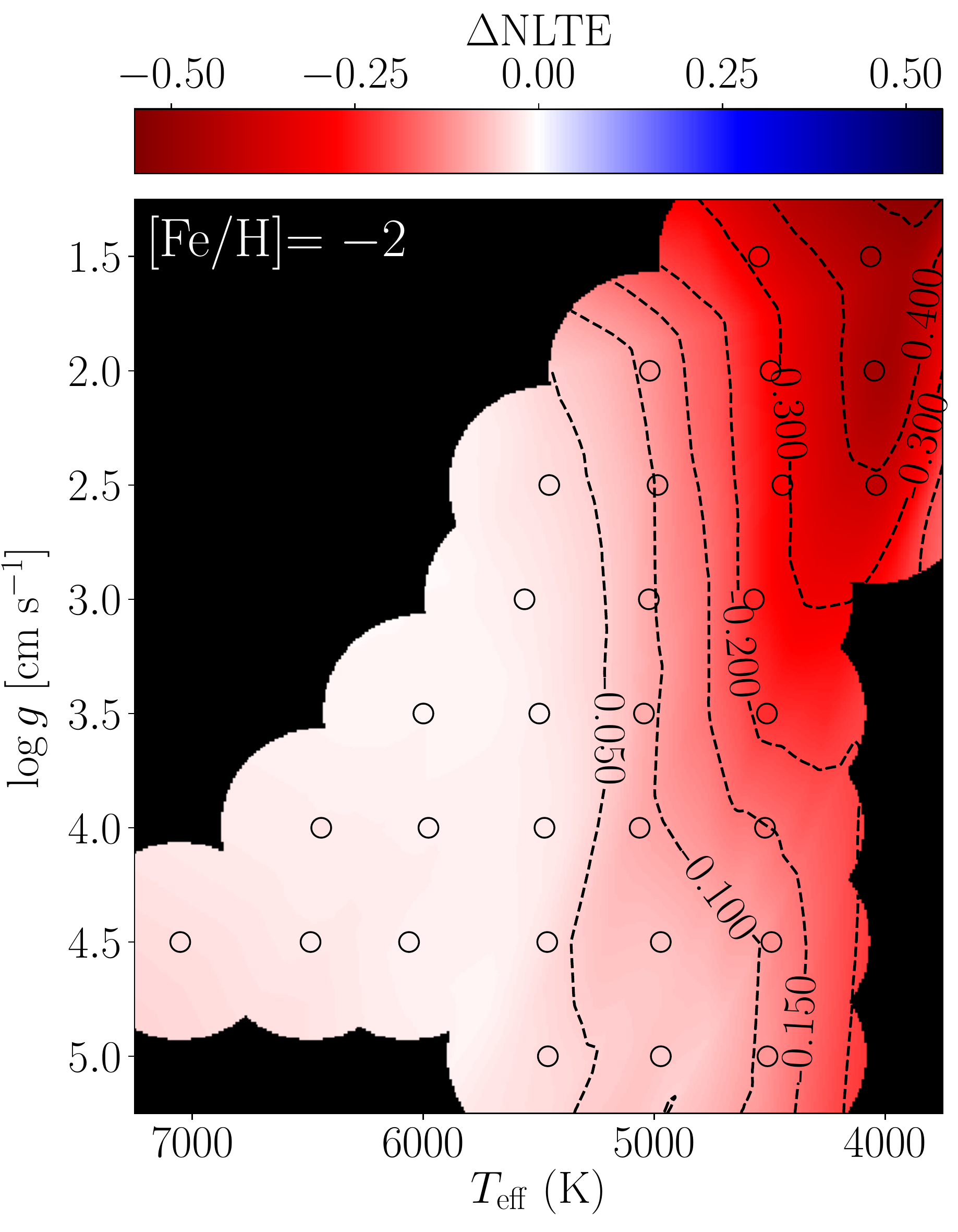}
	\includegraphics[width=0.4\textwidth]{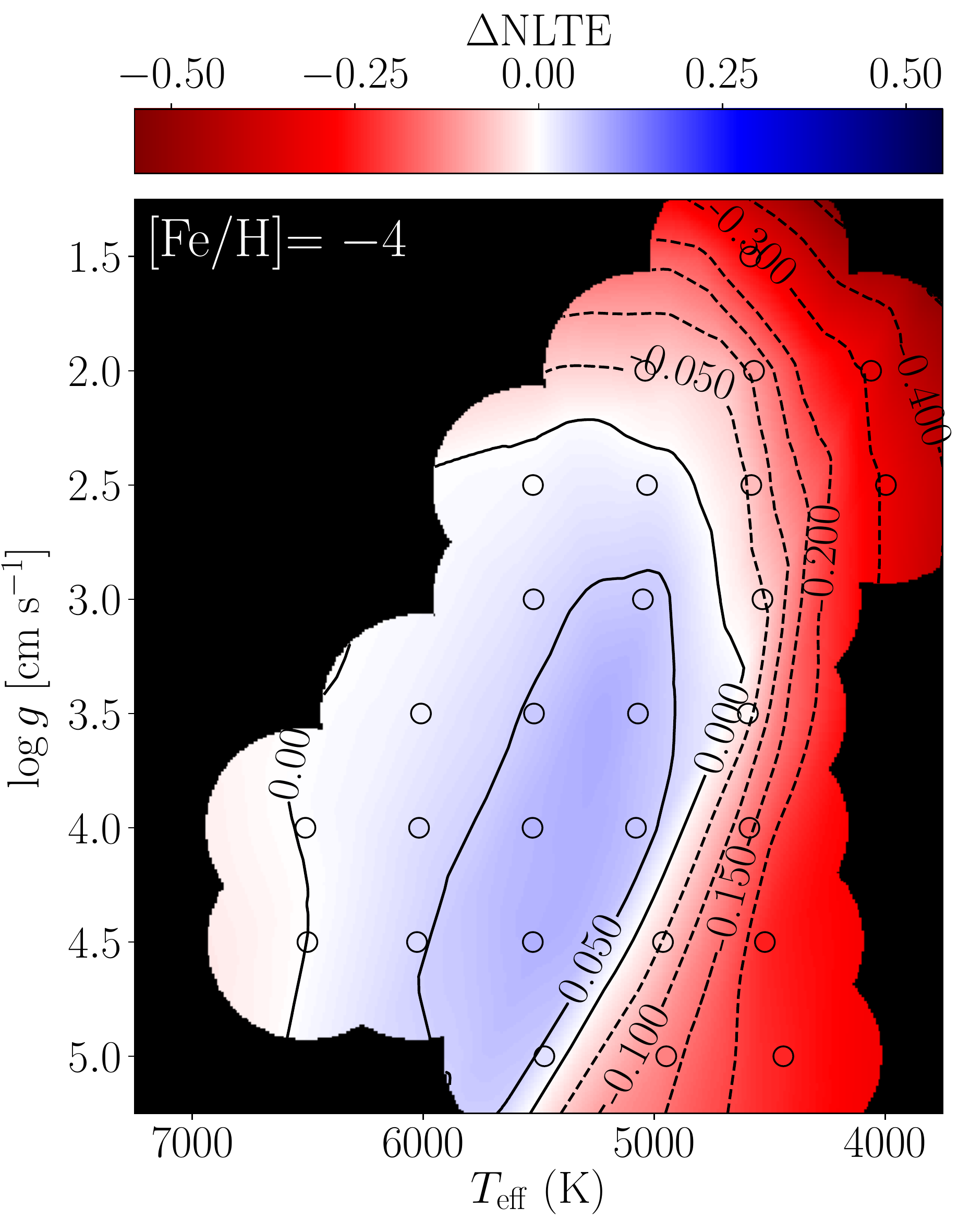}
	\caption{3D NLTE--1D LTE abundance corrections ($\Dn$) for the 670.8\,nm line, shown at a 1D LTE reference abundance of $\ali = 2$. Each panel shows a different metallicity, from left to right, top to bottom, these are: $\met = 0$, $-1$, $-2$, and $-4$.
	The circles show the calculated $\Dn$ from our models, whilst the surface shows the predicted $\Dn$ which has been interpolated according to Section~\ref{sec:int_rew}.
	}
	\label{fig:NLTEc}
\end{figure*}

Fig.~\ref{fig:NLTEc} shows $\Dn$ for the 670.8\,nm line at a representative abundance $\ali = 2$, for four select values of $\met$.
In each panel, we show results for our 3D hydrodynamic models, as well as an interpolation based on a method that is explained further in Section~\ref{sec:int_rew}, to highlight how departures from LTE vary with stellar parameters.

Overall, we find that $\Dn$ tends to become more positive with increasing $\teff$ and decreasing $\met$, but depends only weakly on $\logg$.
At low $\teff$, the line is sufficiently strong that resonance line scattering and over-recombination are both important NLTE effects, together these effects cause the line to increase in strength and therefore yield lower $\ali$.
As $\teff$ increases, the line strength and therefore the effect of resonance line scattering decreases while at the same time overionisation becomes more important. Both effects lead to a weaker line, and thus more positive $\Dn$.
With decreasing line strength, line formation moves inward toward deeper layers, where overionisation becomes less important relative to mutual neutralisation (i.e. recombination) through charge transfer, and the trend therefore turns over when $\teff \sim 6000$\,K in the most metal-poor models.
At higher $\met$, increasing opacities tend to quench overionisation while at the same time higher gas and electron pressure produce higher collisional rates, both of which lead to smaller departures from LTE.
Variations with $\logg$ are very small due to a cancellation effect: the higher pressure leads to increased collisional rates and thus more efficient thermalisation, as well as a decreasing ionisation fraction which strengthens the resonance line and may thus increase the effect of resonance scattering.

We stress that while abundance effects driven by equivalent width differences are a helpful tool to understand departures from LTE and indeed to quickly apply corrections to existing 1D LTE abundance analyses, the line shapes themselves also change. As shown in Fig.~\ref{fig:line_profiles}, while weaker lines often exhibit similar strength in 3D NLTE and 1D LTE leading to small $\Dn$, saturated and strong lines may differ dramatically in both equivalent width and shape.
Using $\Dn$ to derive 3D NLTE $\ali$ may be further hampered by the dependence of 1D LTE analyses on fudge factors such as $\mlt$ and $\vmic$.

\section{Interpolation}
\label{sec:int}
In this section, we show that straightforward spline interpolation of our grid of synthetic spectra yields a larger average error in $\ali$ compared to other interpolation methods.
We therefore investigate more involved interpolation methods, in order to provide to the community a package for accurate abundance analyses based on our synthetic 3D NLTE spectra computed with the \balder\ code: \breidablik\footnote{\url{https://github.com/ellawang44/Breidablik}}.

Interpolation for this grid is complicated due to the fact that $\teff$ is an output rather than input parameter in 3D hydrodynamic simulations. Therefore, the grid is tabulated at irregular intervals in this dimension, with values that are slightly different for every $\logg$ and $\met$. As a result, interpolation methods requiring data points at regular intervals cannot be used out of the box. Additionally, our grid of model atmospheres has relatively large steps in parameter space (see Fig.~\ref{fig:stagger}), resulting in naturally higher errors in interpolation.

There exists many non-linear interpolation methods which can be used to predict abundances from an observed spectrum using an irregular grid. Many of these use inverse modelling, where stellar parameters are predicted from observed spectra rather than the other way around. For example,
\citet{snider01} used artificial neural networks with back-propagation, and applied this to medium resolution spectra to predict stellar parameters;
The Gaia Data Processing and Analysis Consortium uses extremely randomised trees \citep{geurts_extremely_2006}, which is an ensemble method that predicts stellar parameters on the parameter space spanned by the training set \citep{andrae18};
StarNet uses a convolutional neural network applied to APOGEE spectra to predict stellar parameters and 15 elemental abundances in an inverse model \citep{fabbro17};
AstroNN similarly uses a Bayesian neural network with dropout variational inference applied to APOGEE spectra to predict stellar parameters and 18 elemental abundances, with error estimates \citep{leung19}.
The Cannon \bb{\citep{ness15, ho16, casey16}} uses a polynomial model to determine stellar parameters and elemental abundances from a spectrum.
However unlike the other models, \bb{the} Cannon can also be applied to forward modelling, where it generates a spectrum given stellar parameters and elemental abundances\bb{.}
Another forward modelling approach is the Payne \bb{\citep{ting19}}, which uses a fully connected feedforward neural network to rapidly and accurately predict spectra given stellar parameters and abundances, followed by a fit to APOGEE spectra to determine stellar parameters and 20 elemental abundances.

While the inverse models have certain advantages, e.g. in terms of speed and always returning an answer, forward modelling is more flexible.
For \breidablik, we therefore choose a forward modelling approach, and provide tools that fit our synthetic spectra to observations.
This has the advantage that our method is not tied to any particular instrumental setup, but can be implemented as part of any traditional analysis. It can also be further integrated into more advanced methodologies, that evaluate systematic errors or advanced statistics including deriving upper limits to non-detections.

We investigate the interpolation of both spectral line profiles and line strengths.
The interpolation of spectral line profiles is subtly different from that of line strengths as they involve the extra dimension of wavelengths.
While line strengths vary smoothly across stellar parameter space, their \textit{shape} is more complicated due to broadening processes and hydrodynamic velocity fields that may redistribute flux between adjacent wavelength points in a nonlinear way.

We test a number of different interpolation methods for both line profiles and strengths, and evaluate them using leave-one-out cross-validation.
Hyperparameters for each method are optimised through 5-fold cross-validation, as per standard practice (see \citealp[ch.5]{james}; \citealp[ch.4]{kuhn}), using a common seed to generate the folds.
As spectrum variations across stellar parameters are more difficult to predict than variations with $\ali$, folds were created based on $\teff$, $\logg$, and $\met$ only.
The final model is trained using the regular models in the \stagger\ grid, and validated on models tailored to particular stars, that were not part of the training set. Sections~\ref{sec:int_prof} and \ref{sec:int_rew} discuss interpolating spectral line profiles and line strengths respectively, presenting and comparing the different interpolation methods tested.

\subsection{Interpolating Spectral Line Profiles}
\label{sec:int_prof}
We test and compare three interpolation methods: spline interpolation, the Cannon, and Kriging. Our implementation of the different methods and the hyperparameters used are described in sections~\ref{sec:tri}-\ref{sec:kri}.

\begin{figure}
	\centering
	\includegraphics[width=0.49\textwidth]{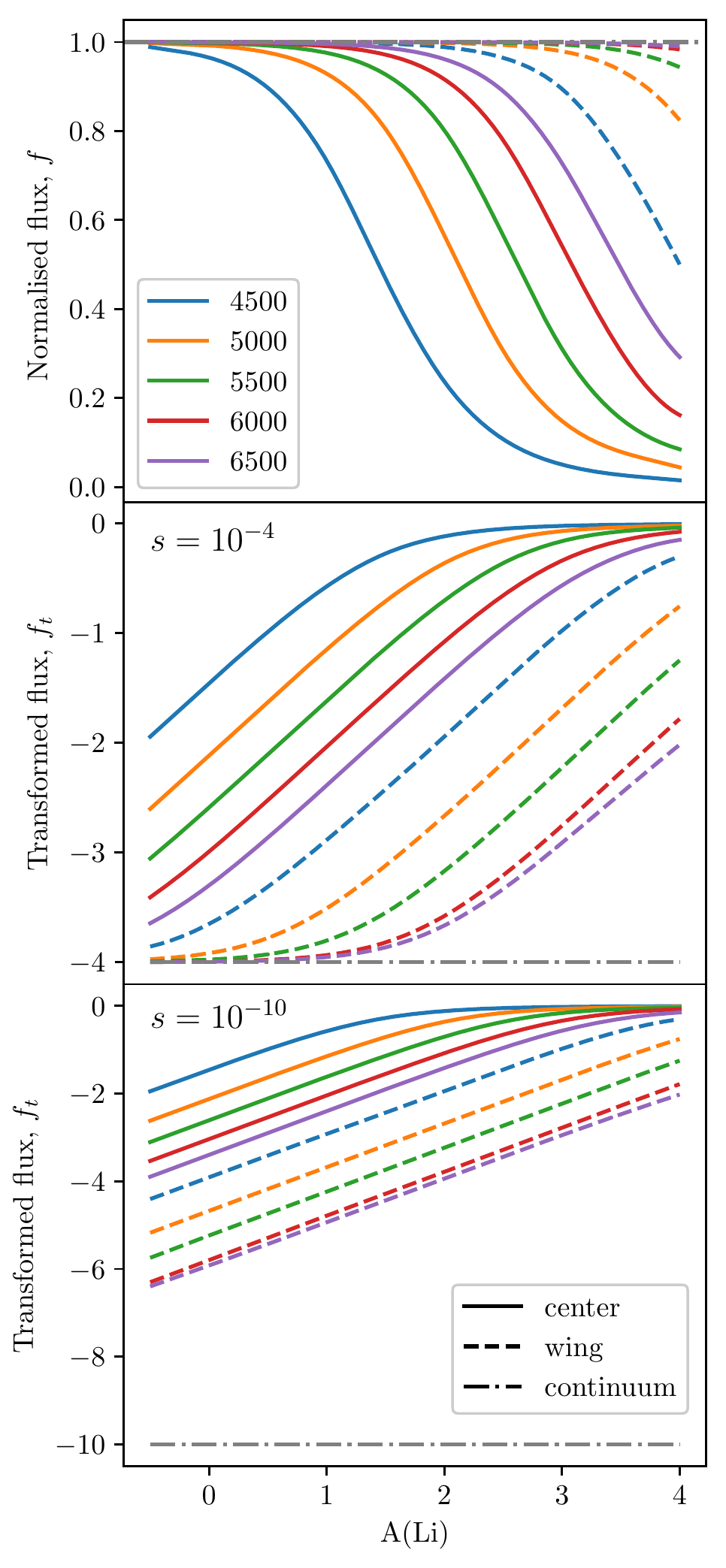}
	\caption{Comparison of the normalised flux and how it varies as a function of $\ali$ (top panel), to the transformed flux $f_t$ (middle and bottom panels) for two different values of the softening parameter $s$.
	Solid lines show the flux at line centre, dashed lines represent a point in the line wing (offset by $-0.03$\,nm from the core).
	and the dash-dotted line is the continuum. Models of different $\teff$ are shown with different colour; all models use $\logg = 4$ and $\met = -2$.}
	\label{fig:flux_transform}
\end{figure}

In order to make the normalised flux easier to interpolate, especially for very weak lines, we transform it to a quantity that scales better with abundance. On the weak part of the curve of growth, the flux depression varies approximately linearly with the number of absorbers, so we define a transformed flux
\begin{equation}
f_t = \log_{10}(1 - f + s),
\label{eq:transform}
\end{equation}
where
$f$ is the normalised flux, and $s$ is a small positive constant (treated as a hyperparameter) which smoothly truncates $f_t$ as the flux approaches the continuum.
Fig.~\ref{fig:flux_transform} illustrates how the normalised flux compares to the quantity $f_t$, for two representative values of $s$.
Importantly, $f_t$ varies nearly linearly with $\ali$ for small flux depressions, with a dynamic range controlled by $s$, before reaching saturation as the flux approaches zero. Conversely, as the flux approaches the continuum we find $f_t = \log_{10} s$, which helps avoid numerical noise dominating the model.
Test runs where we interpolate over the normalised flux $f$ tend to incur larger relative errors when the flux depression is small, corresponding to larger abundance errors and larger errors in the detailed shape for very weak lines.

All interpolation methods are trained on all 3 Li lines and on a per pixel basis.

\subsubsection{Spline Interpolation}
\label{sec:tri}
Local polynomial models like spline interpolation are the most straightforward, and most commonly used in the literature.
Due to the irregular spacing of $\teff$ in our grid of 3D models, we perform a ``training'' step following \citet{amarsi18}, wherein we create a finer grid with regular steps in stellar parameters $\teff$, $\logg$ and $\met$ using cubic spline interpolation and linear extrapolation.
We then use trilinear interpolation within this finer grid to interpolate to the exact stellar parameters at each $\ali$, and produce the final spectrum using cubic splines over $\ali$.
This method uses only a single hyperparameter, which we optimise to $s = 10^{-9}$ (see Eq.~\ref{eq:transform}).

\subsubsection{The Cannon}
\label{sec:can}
The Cannon is a global polynomial interpolation method \citep{ness15, ho16}. We provide the Cannon with 3 labels: $\teff$, $\logg$, and $\met$, training one model per $\ali$, then using cubic spline over $\ali$ to compute the final interpolated line profile. The Cannon is normally trained by applying a $\chi^2$ minimisation method to observed data. Since our synthetic data are not stochastic, it therefore does not carry statistical uncertainties, and as such, we set the variance in pixel to a constant value.
Our hyperparameter test finds best results when using a third order polynomial form with 20 terms (including cross terms) describing the flux in each pixel, and $s = 10^{-9}$.

\subsubsection{Kriging}
\label{sec:kri}
Kriging interpolates by computing the distance dependent weighted averages of points in a neighbourhood. We define this neighbourhood as the entire grid to remove discontinuities. We use the \texttt{OrdinaryKriging3D} class from the Python package \texttt{PyKrige} to interpolate across stellar parameters individually at each $\ali$, and then interpolate over $\ali$ by cubic splines to produce the final spectrum. Our best hyperparameter is: $s = 10^{-7}$.

\subsubsection{Errors: 3D leave-one-out}
\label{sec:3Dloo}
We summarise the abundance fitting statistics over the 670.8\,nm line for the three investigated models in Table~\ref{table:errors}.
Our error statistics are based on the MAD (median absolute deviation) and RMS (root mean square error) measures.

\begin{table*}
\begin{tabular}{c|cccccccc}
& \multicolumn{2}{c}{3D leave-one-out} & \multicolumn{2}{c}{1D interpolation} & \multicolumn{2}{c}{1D extrapolation} \\
Method & MAD & RMS & MAD & RMS & MAD & RMS & training time (s) & execution time (s) \\
\hline
Spline & 0.046 & 0.115 & 0.021 & 0.060 & 0.028 & 0.219 & 1880 & 0.0941 \\
The Cannon & 0.020 & 0.029 & 0.085 & 0.176 & 0.062 & 0.088 & 11.1 & 0.138 \\
Kriging & 0.014 & 0.022 & 0.020 & 0.038 & 0.019 & 0.143 & 235 & 5.62 \\
\hline
\end{tabular}
\caption{Comparison of different interpolation methods used to predict the 670.8\,nm line profile, showing their MAD and RMS error statistics in predicting abundance for leave-one-out cross-validation on the 3D grid, and interpolation and extrapolation tests on a 1D grid. Also shown is the time taken to train the models for all three Li lines (418 pixels total), and the time it takes to execute the model to produce interpolated line profiles for all three lines with arbitrary values of $\teff$, $\logg$, $\met$, and $\ali$.}
\label{table:errors}
\end{table*}

For spline interpolation, we find that models located near the edges of the grid incur very large errors that drive a large RMS error statistic.
This is the expected behaviour for spline models, as they tend to extrapolate poorly. A particular difficulty is the fact that our grid does not have regular edges, but the upper and lower limits on $\teff$ vary with $\logg$.
In contrast, we find that both The Cannon and Kriging performs very well on average, and are relatively reliable also when extrapolating to edge models.
We select the interpolation method with the smallest MAD error statistic, Kriging, to use in \breidablik. We opt for the MAD statistic rather than the RMS error as the former is less sensitive to outliers, and more representative of the typical expected error.

Table~\ref{table:errors} shows also representative values for the training time and execution time for each interpolation method, as executed on a single CPU core on a modern desktop computer.
The training time represents the execution time required to determine the model with our optimum set of hyperparameters, and
the execution time is the time required to predict the lithium abundance for a single spectral line.
The Cannon is by far the fastest in terms of training time due to its use of weighted least squares fitting and its simple polynomial form. In comparison, the training time of Kriging and spline interpolation methods are slower by several orders of magnitude.

\subsubsection{Errors: Verification Models}
\label{sec:verificationmodels}
\begin{figure*}
	\centering
	\includegraphics[width=0.485\textwidth]{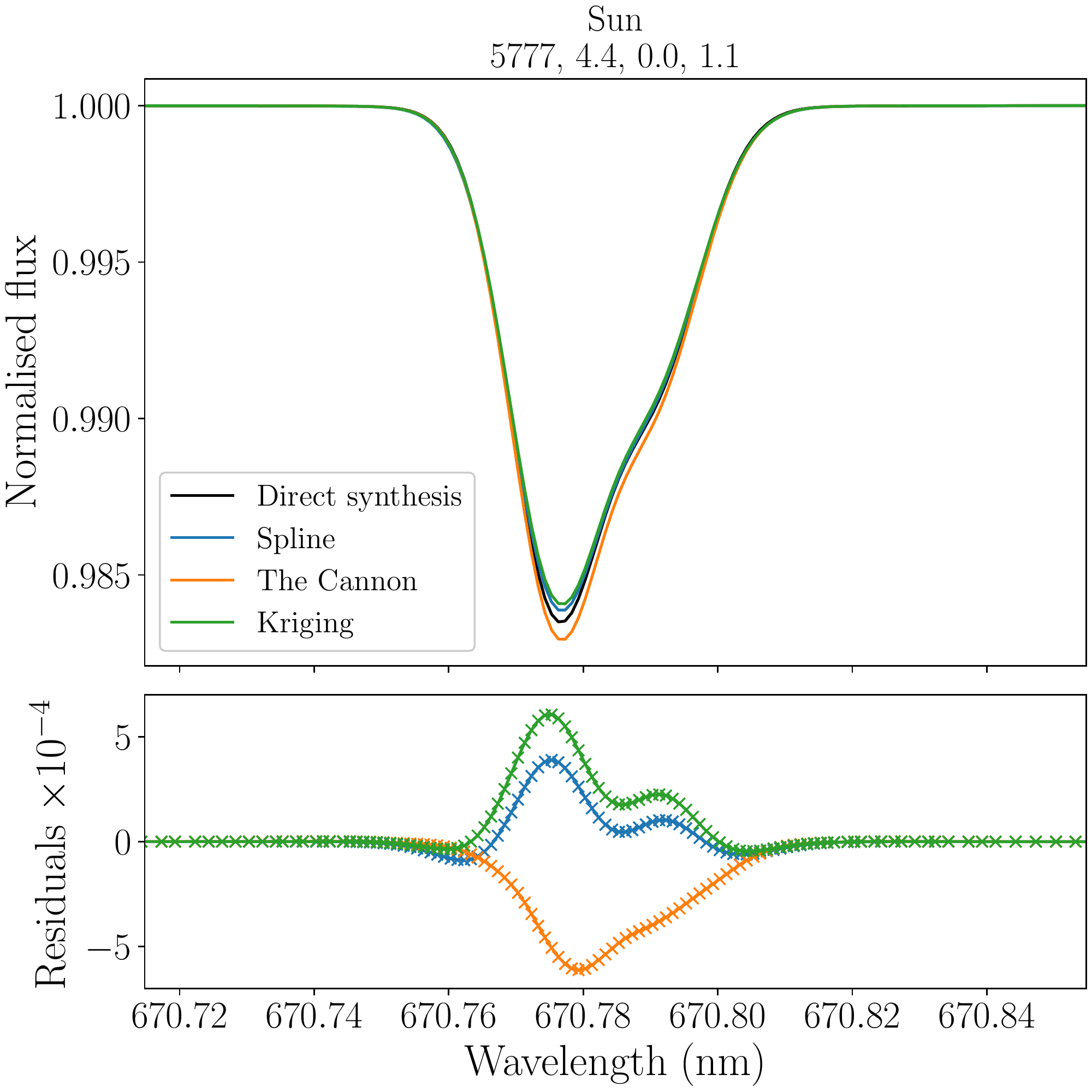}
	\includegraphics[width=0.485\textwidth]{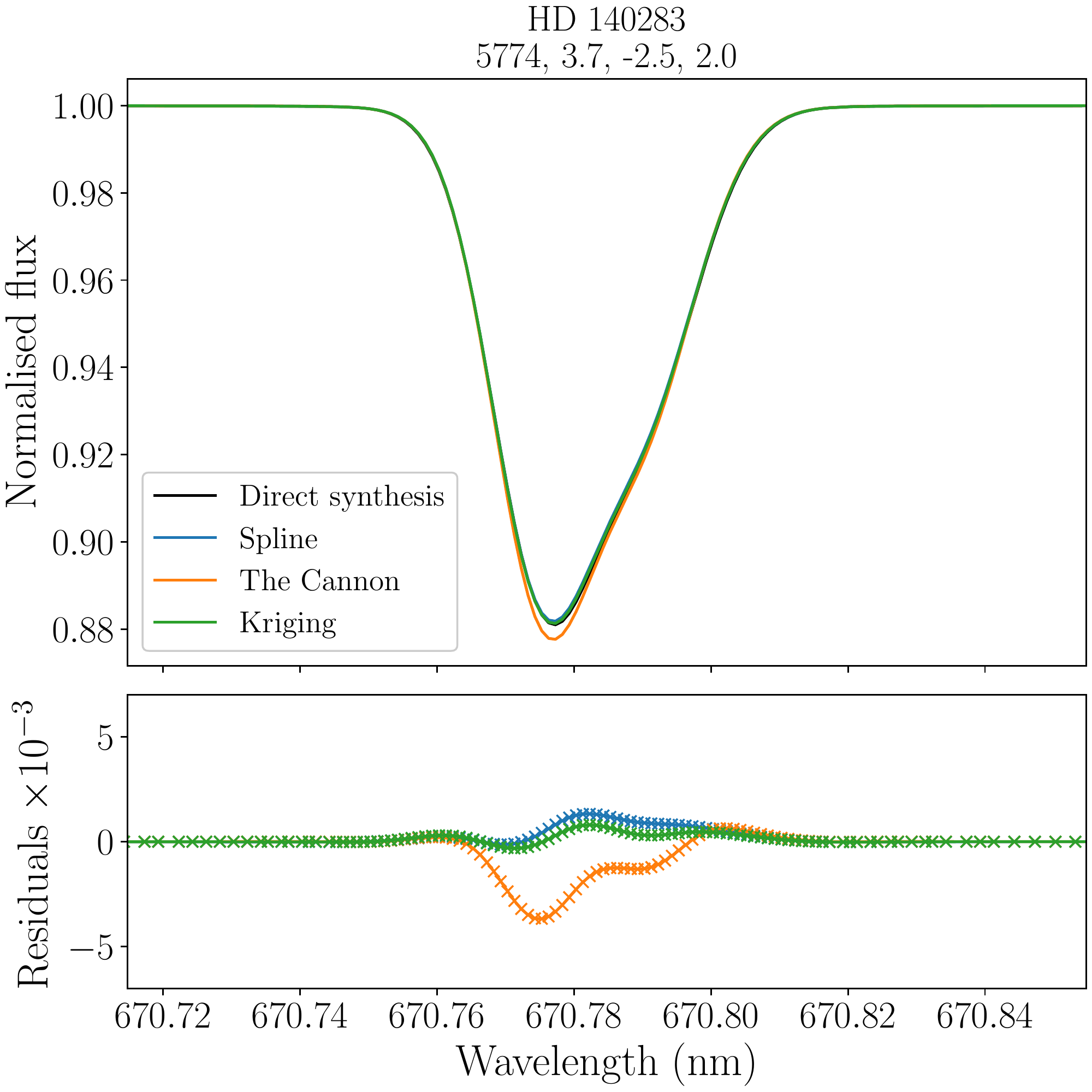}
	\includegraphics[width=0.485\textwidth]{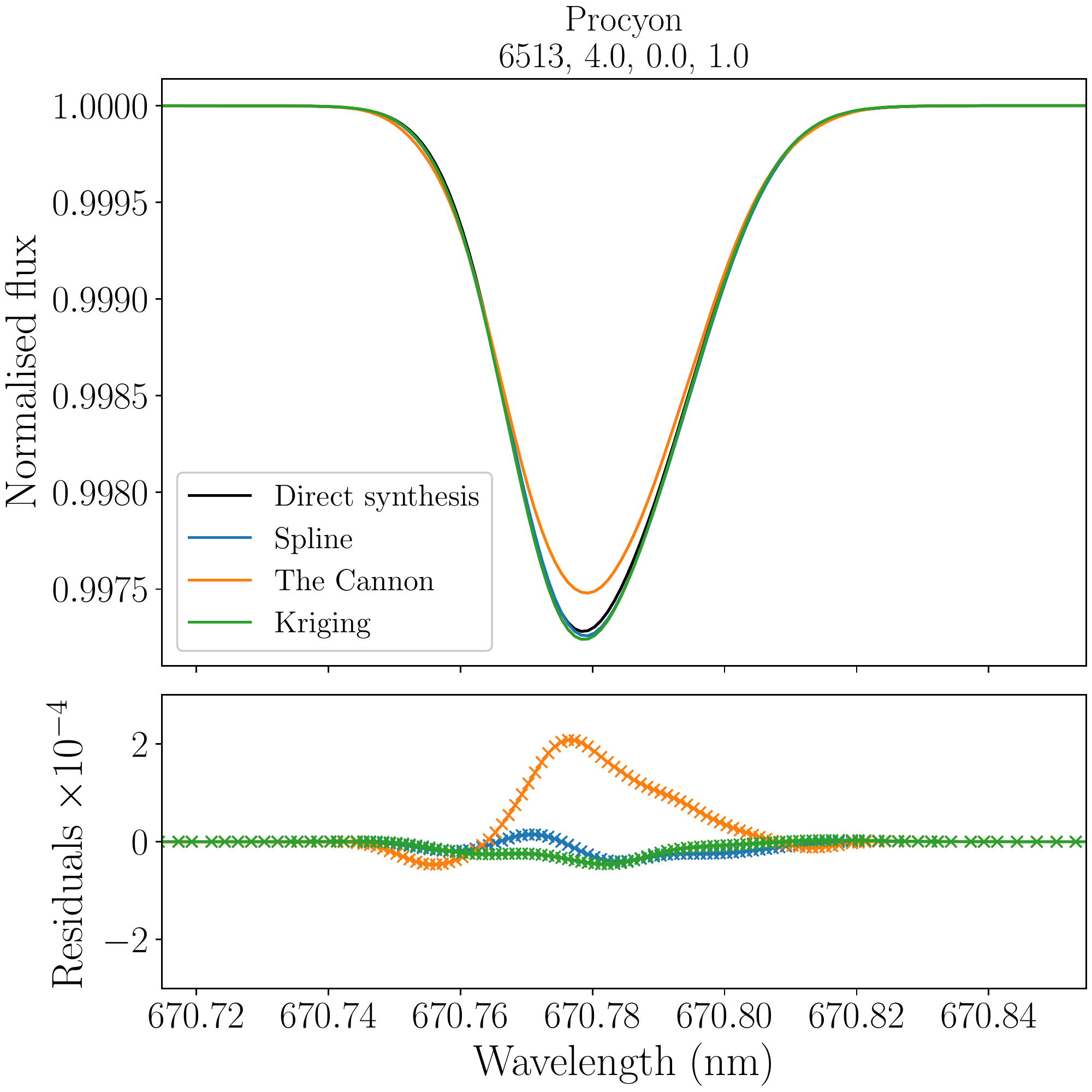}
	\includegraphics[width=0.485\textwidth]{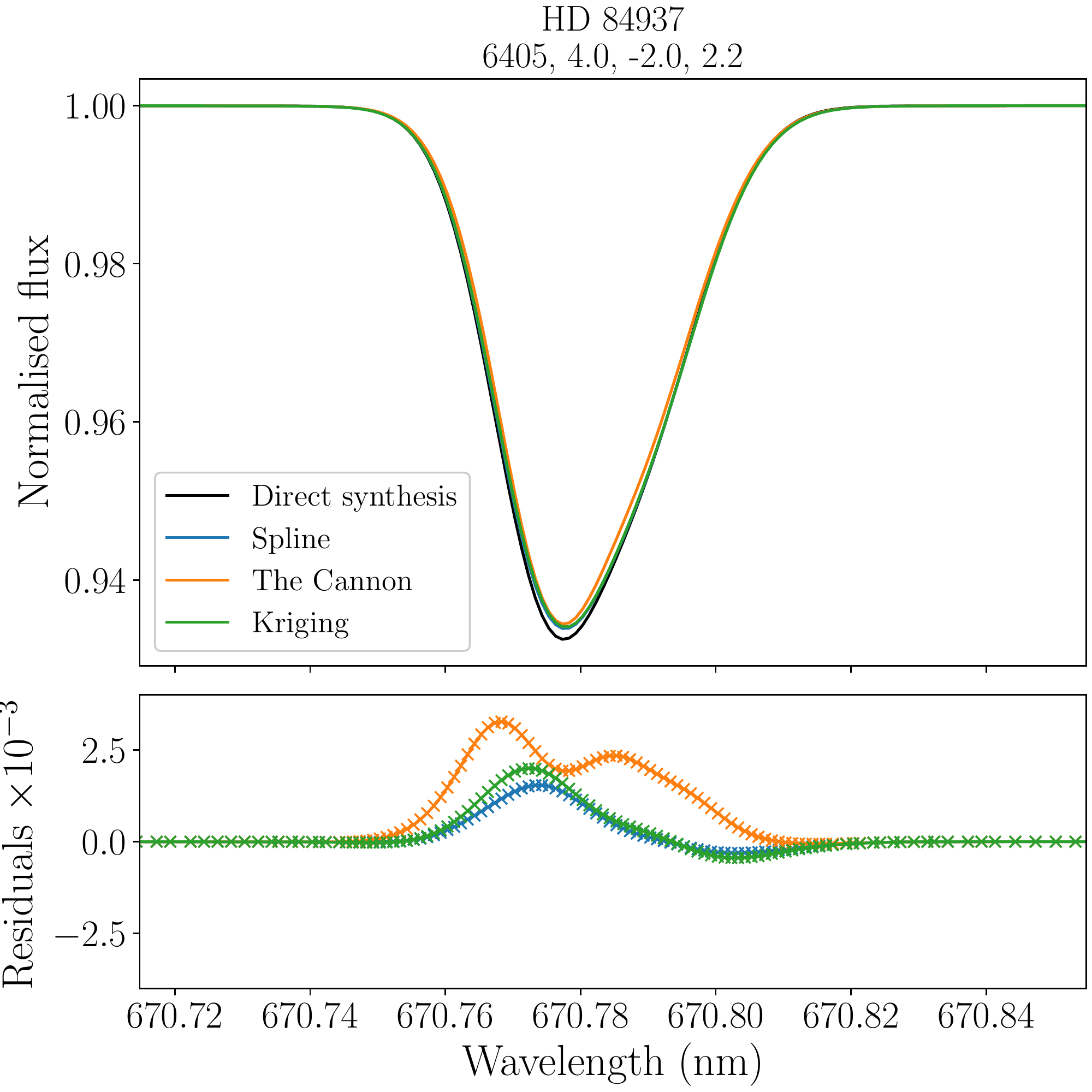}
	\caption{Comparison of interpolation models with verification models (that were not trained on). The verification models were tailored to specific stars, indicated by name and stellar parameters: $\teff$, $\logg$, $\met$, and $\ali$. Note that the residuals between prediction and verification model have been magnified by a different amount in each plot.
    }
	\label{fig:ver}
\end{figure*}

To verify the validity of the final interpolation models, and to show the errors expected in practice, we used a verification set of four models that are not part of the regular \stagger\ grid but rather were tailored to individual stars.
The exact line profiles computed with \balder\ for these verification models are compared to the interpolated line profiles from the three tested interpolation methods in Fig.~\ref{fig:ver}, and the measured $\ali$ using each interpolation method is reported in Table~\ref{table:ver_prof}.
The interpolated line profiles have a similar shape to the synthesised profiles and replicates the characteristic asymmetry of the Li line.
Overall, we find that spline interpolation and Kriging both perform well.
Whilst the Cannon has been used successfully in many different cases, mainly in applications to large sets of empirical spectra \citep{ness15, ho_label_2017, buder_galah_2018}, we note that the Cannon does not perform as well as other interpolation methods for our particular data set due to the large span in stellar parameter space.
Compared to Table~\ref{table:errors}, all verification model errors are notably smaller than the measured leave-one-out errors. As all four verification models are located either near regular models or well inside the grid boundaries, spline and Kriging perform much better on the verification models compared to the measured MAD error statistic in the leave-one-out tests.
These verification results indicate that the final interpolation methods produce consistent results when compared to the leave-one-out cross-validation errors.

\begin{table*}
\begin{tabular}{c|cccc}
\diagbox{Method}{Model} & Sun & HD 140283 & Procyon & HD 84937 \\
\hline
Direct synthesis & 1.100 & 2.000 & 1.000 & 2.200 \\
\hline
Spline & 0.006 & 0.004 & -0.004 & 0.007 \\
The Cannon & -0.015 & -0.010 & 0.026 & 0.019 \\
Kriging & 0.012 & 0.002 & -0.006 & 0.009 \\
\hline
\end{tabular}
\caption{Error in the predicted $\ali$ using line profile interpolation models on verification models (that were not trained on). The input $\ali$ used is shown in the first row, labeled `direct synthesis', while subsequent rows show the error relative to the reference. The verification models were tailored to specific stars, indicated by name, whose stellar parameters are given in Fig.~\ref{fig:ver}.}
\label{table:ver_prof}
\end{table*}

\subsubsection{Errors: 1D comprehensive grid}
\label{sec:1Dcomp_grid}
Leave-one-out cross-validation errors presented in Section~\ref{sec:3Dloo} likely yields overestimated error estimates. This is because the effective interpolation step in these tests will be 500\,K in $\teff$, 0.5\,dex in $\logg$ and 0.5--1\,dex in $\met$, while in practice interpolation on the full grid will require steps of at most 250\,K in $\teff$, 0.25\,dex in $\logg$ and 0.25--0.5\,dex in $\met$.
While the tests on verification models presented in Section~\ref{sec:verificationmodels} yield fair estimates of the interpolation error, they cover only a very limited parameter space.
Due to the large computational cost of creating 3D hydrodynamic simulations, it is beyond our computational resources to generate additional verification models that cover the entire parameter space for comparison to our 3D NLTE calculations.
In this section, we present errors from a more comprehensive grid of verification models based on 1D NLTE calculations.

We use spectra computed using 1D hydrostatic models that were computed with the \atmo\ code, adopting $\mlt = 1.5$, $\vmic = 1$\,\kms. We train spectrum interpolation models using the same hyperparameters as were used for our 3D NLTE spectra.
We test on a total of 2000 1D \atmo\ models with randomly generated values for $\teff$ and $\logg$, while discrete values of $\met$ are selected from the existing grid, and $\ali$ is drawn from a normal distribution with mean $2.5$ and standard deviation $0.3$. 1000 of these models are bounded by the existing grid models, hence are ``interpolation'' models; the other 1000 models are not fully bounded by existing grid models, hence are ``extrapolation'' models.

The MAD and RMS errors determined for these two samples are shown in Table~\ref{table:errors}.
We find that spline interpolation performs well on average on both interpolation and extrapolation, while a small number of extreme outliers in the extrapolation sample produce a large RMS error.
The Cannon surprisingly performs better on extrapolation than interpolation, indicating that there are likely small oscillations in the model. Our preferred model, Kriging, performs well on both interpolation and extrapolation, again with a small number of extreme outliers producing a large RMS error in the extrapolation sample.

\subsection{Interpolating Line Strengths}
\label{sec:int_rew}
As a complement to our interpolation of spectral line profiles, we also provide Python classes in \breidablik\ to derive abundances and abundance corrections based on reduced equivalent widths for all 3 Li lines.
Our method for providing abundance corrections, $\Dn$, takes as input stellar parameters and $\ali$ measured in 1D LTE.
Our method for providing 3D NLTE abundance measurements, \bb{$\textrm{A}_{\textrm{REW}}$}, takes as input stellar parameters and a REW (Eq.~\ref{eq:REW}).

We test a total of five interpolation models, including the spline interpolation, the Cannon and Kriging, presented in Sections~\ref{sec:tri} - \ref{sec:kri}, which are used here with no need for hyperparameters because the flux transform is not used.
We introduce here two additional interpolation methods: Multi-Layer Perceptron (MLP) and Support Vector Regression (SVR).
These methods were also evaluated in our tests for spectral line interpolation, but we found that due to limited precision and the random nature of the training procedure, both methods produce pixel-to-pixel variations that resemble noise in the line profiles. As a result, MLP and SVR are not considered for interpolating line profiles.

We note that \citet{mott20} present a method similar to our application of the Cannon, where they use a polynomial fit over $\teff$, $\logg$ and $\ali$, and derive fitting functions individually at each value of $\met$. They test their fitting functions on their input data, and find that errors are typically in the range of 0.020\,dex for both $\Dn$ and \bb{$\textrm{A}_{\textrm{REW}}$}.

\subsubsection{Support Vector Regression}
Support Vector Regression (SVRs) is a form of Support Vector Machines \citep{vapnik95},
that defines a subset of the training set (the support vectors) which fall within $\varepsilon$ of the geometric margin separating data, using a particular kernel function to transform the input space.
We use the \texttt{SVR} class \citep{chang11} from the Python package \textit{scikit-learn} \citep{scikit-learn}. For $\Dn$, we find best hyperparameters: the penalty parameter, $C = 100$; penalty distance, $\varepsilon = 10^{-3}$; and kernel function, $f_{\rm{ker}} =$ radial basis function (rbf). For \bb{$\textrm{A}_{\textrm{REW}}$}, we find best hyperparameters: $C = 5000$, $\varepsilon = 10^{-5}$, and $f_{\rm{ker}} = $ rbf.

\subsubsection{Multi-layer Perceptron}
\label{sec:mlp}
Multi-layer Perceptrons (MLPs) are fully connected feed forward neural networks, that connect ``neurons'' in a series of layers through a mixture of linear and non-linear transforms. We use the \texttt{MLPRegressor} class \citep{hinton90} from the Python package \textit{scikit-learn} \citep{scikit-learn}. We set the maximum number of iterations to $10^5$ and the tolerance to $10^{-6}$. For $\Dn$, our best hyperparameters are: number of layers, $n_l = 2$; the number of neurons per layer, $n = 900$; the L2 penalty, $\alpha = 0.1$; and the activation function, $f_{\rm{act}} =$ rectified linear unit (ReLU). For \bb{$\textrm{A}_{\textrm{REW}}$}, the best hyperparameters are: $n_l = 2$, $n = 350$, $\alpha = 0.01$, and $f_{\rm{act}} =$ ReLU.

\subsubsection{Errors}
We show the fitting statistics
MAD and RMS errors of $\Dn$ and \bb{$\textrm{A}_{\textrm{REW}}$} models in Table~\ref{table:line_strengths}. These reported errors are also likely to be upper bounds on the errors expected from fully trained grids, similar to Table~\ref{table:errors}.
SVR takes much longer to train over \bb{$\textrm{A}_{\textrm{REW}}$} compared to $\Dn$ because the training time is sensitive to the hyperparameter, $C$, where larger values of $C$ tends to takes longer to train due to higher numbers of iterations required.
MLP takes longer to train over $\Dn$ compared to \bb{$\textrm{A}_{\textrm{REW}}$}, because the training time is sensitive to $n$, as larger $n$ implies a higher number of fitted parameters.
Overall, our preferred method, MLP, produces by far the lowest errors and executes rapidly.

\begin{table*}
\begin{tabular}{cc|cccccccc}
& & \multicolumn{2}{c}{3D leave-one-out} & \multicolumn{2}{c}{1D interpolation} & \multicolumn{2}{c}{1D extrapolation} & & \\
Line strengths & Interpolation & MAD & RMS & MAD & RMS & MAD & RMS & training time [s] & execution time [ms] \\
\hline
\multirow{5}{5em}{$\Dn$} & Spline & 0.040 & 0.083 & 0.012 & 0.139 & 0.020 & 0.209 & 29.3 & 0.578 \\
& The Cannon & 0.025 & 0.041 & 0.024 & 0.148 & 0.027 & 0.199 & 0.198 & 0.00263 \\
& Kriging & 0.021 & 0.037 & 0.012 & 0.142 & 0.012 & 0.170 & 0.567 & 15.1 \\
& SVR & 0.017 & 0.027 & 0.018 & 0.140 & 0.014 & 0.163 & 27.1 & 0.119 \\
& MLP & 0.012 & 0.020 & 0.012 & 0.140 & 0.011 & 0.173 & 2140 & 0.384 \\
\hline
\multirow{5}{5em}{\bb{$\textrm{A}_{\textrm{REW}}$}} & Spline & 0.051 & 0.663 & 0.026 & 0.063 & 0.024 & 0.202 & 28.3 & 0.563 \\
& The Cannon & 0.038 & 0.066 & 0.026 & 0.057 & 0.020 & 0.069 & 0.273 & 0.00209 \\
& Kriging & 0.013 & 0.022 & 0.025 & 0.046 & 0.018 & 0.149 & 0.552 & 14.9 \\
& SVR & 0.018 & 0.036 & 0.030 & 0.056 & 0.034 & 0.080 & 1270 & 0.170 \\
& MLP & 0.010 & 0.014 & 0.027 & 0.047 & 0.015 & 0.073 & 133 & 0.176 \\
\hline
\end{tabular}
\caption{Comparison of different interpolation methods used to predict the abundance correction ($\Dn$) and \bb{abundance based on equivalent widths ($\textrm{A}_{\textrm{REW}}$)}, for the 670.8\,nm line. The columns are the same as in Table~\ref{table:errors}.}
\label{table:line_strengths}
\end{table*}

To verify the validity of the final line strength interpolation models, we use the same verification set as in Section~\ref{sec:int_prof}. The exact $\Dn$ and 3D NLTE $\ali$ computed with \balder\ for these verification models are compared to the interpolated line strengths from the five tested methods in Table~\ref{table:ver_str}. The verification results for the Cannon, Kriging, SVR, and MLP in Table~\ref{table:ver_str} mostly matches the results in Table~\ref{table:line_strengths}. Spline interpolation and Kriging both perform significantly better for the verification models again due to the location of the verification models being near regular models or well inside grid boundaries.
These verification results indicate that the final interpolation models for both $\Dn$ and \bb{$\textrm{A}_{\textrm{REW}}$}
are consistent with the leave-one-out cross-validation error statistics.

\begin{table*}
\begin{tabular}{cc|cccc}
Line strengths & \diagbox{Method}{Model} & Sun & HD 140283 & Procyon & HD 84937 \\
\hline
\multirow{6}{5em}{$\Dn$} & Direct synthesis & -0.065 & -0.006 & 0.040 & -0.026 \\
\cline{2-6}
& Spline & 0.012 & 0.021 & -0.036 & -0.004 \\
& The Cannon & 0.000 & 0.015 & -0.045 & 0.005 \\
& Kriging & 0.045 & 0.016 & 0.053 & -0.001 \\
& SVR & 0.007 & 0.026 & -0.029 & 0.005 \\
& MLP & 0.006 & 0.019 & -0.041 & 0.005 \\
\hline
\multirow{6}{5em}{\bb{$\textrm{A}_{\textrm{REW}}$}} & Direct synthesis & 1.100 & 2.000 & 1.000 & 2.200 \\
\cline{2-6}
& Spline & 0.000 & 0.000 & -0.004 & 0.005 \\
& The Cannon & -0.017 & -0.006 & 0.026 & 0.016 \\
& Kriging & 0.011 & 0.002 & -0.005 & 0.006 \\
& SVR & 0.009 & -0.008 & -0.018 & 0.008 \\
& MLP & 0.002 & -0.002 & -0.008 & 0.008 \\
\hline
\end{tabular}
\caption{Error in the predicted abundance correction ($\Dn$) and \bb{abundance based on equivalent widths ($\textrm{A}_{\textrm{REW}}$)} for verification models (that were not trained on), based on equivalent widths for the 670.8\,nm line. The actual values of $\Dn$ and 3D NLTE $\ali$ computed for each verification model is shown in the first row labeled `direct synthesis', while subsequent rows show the errors, as in Table~\ref{table:ver_prof}.
}
\label{table:ver_str}
\end{table*}

We also compute errors for the $\Dn$ and \bb{$\textrm{A}_{\textrm{REW}}$} models on a 1D comprehensive grid, similar to Section~\ref{sec:1Dcomp_grid}. Since these models are trained on 1D hydrostatic model atmospheres, the $\Dn$ model takes as input the stellar parameters and 1D LTE $\ali$ and returns $\dn =$ 1D NLTE $\ali$ - 1D LTE $\ali$; whilst the \bb{$\textrm{A}_{\textrm{REW}}$} model takes stellar parameters and REW and outputs 1D NLTE $\ali$. The MAD and RMS errors for these 1000 ``interpolation'' and 1000 ``extrapolation'' models are shown in Table~\ref{table:line_strengths}. All models perform well on average for both interpolation and extrapolation, but extrapolation tends to have more extreme outliers, as seen from the higher RMS for extrapolation. The $\Dn$ models have comparable MAD and RMS errors between the 3D leave-one-out cross-validation errors and 1D interpolation and extrapolation errors, with the exceptions of spline and the Cannon. Spline and Kriging for the \bb{$\textrm{A}_{\textrm{REW}}$} models have more extreme outliers when extrapolating compared to interpolating; whilst the Cannon, SVR, and MLP have almost comparable RMS, indicating that they perform better extrapolating compared to other methods. Overall, our preferred model, MLP, performs well in both interpolation and extrapolation compared to other models.

\citet{mott20} presented polynomial fits with typical errors of 0.020\,dex in both $\Dn$ and 3D NLTE REW. We note that although our errors are similar in magnitude, these errors are not directly comparable. This is because the parameters range in the \citet{mott20} grid is smaller than ours, and the tests used to derive the error statistics are also different. We believe a leave-one-out cross-validation and an additional test on randomly selected models should be more representative of errors typically seen in real data sets.

In addition, we use MLP to fit the 610.4\,nm and 812.6\,nm line. The hyperparameters, leave-one-out cross-validation errors, and verification results are provided in Appendix~\ref{app:rew_other}.

\section{Discussion}
\label{sec:dis}
As shown in Section~\ref{sec:UV}, our NLTE results are quantitatively different from previous literature, in particular resulting in significantly more negative abundance corrections.
Ordinarily, adjustments to abundance scales like this are tested through the agreement of different abundance diagnostics in standard stars. Due to the very low abundance of lithium, only the resonance line at 670.8\,nm is visible in most stars, and indeed even this line is difficult to measure in the Sun where the photospheric abundance is depleted by more than 2\,dex relative to the protosolar value \citep[e.g.][]{asplund09}.
While the subordinate line at 610.4\,nm is inaccessible in the solar spectrum, it is sometimes possible to measure in spectra of exceptional quality or when the lithium abundance is strongly enhanced (typically causing the 670.8\,nm line to saturate).

\begin{figure}
	\includegraphics[width=0.49\textwidth]{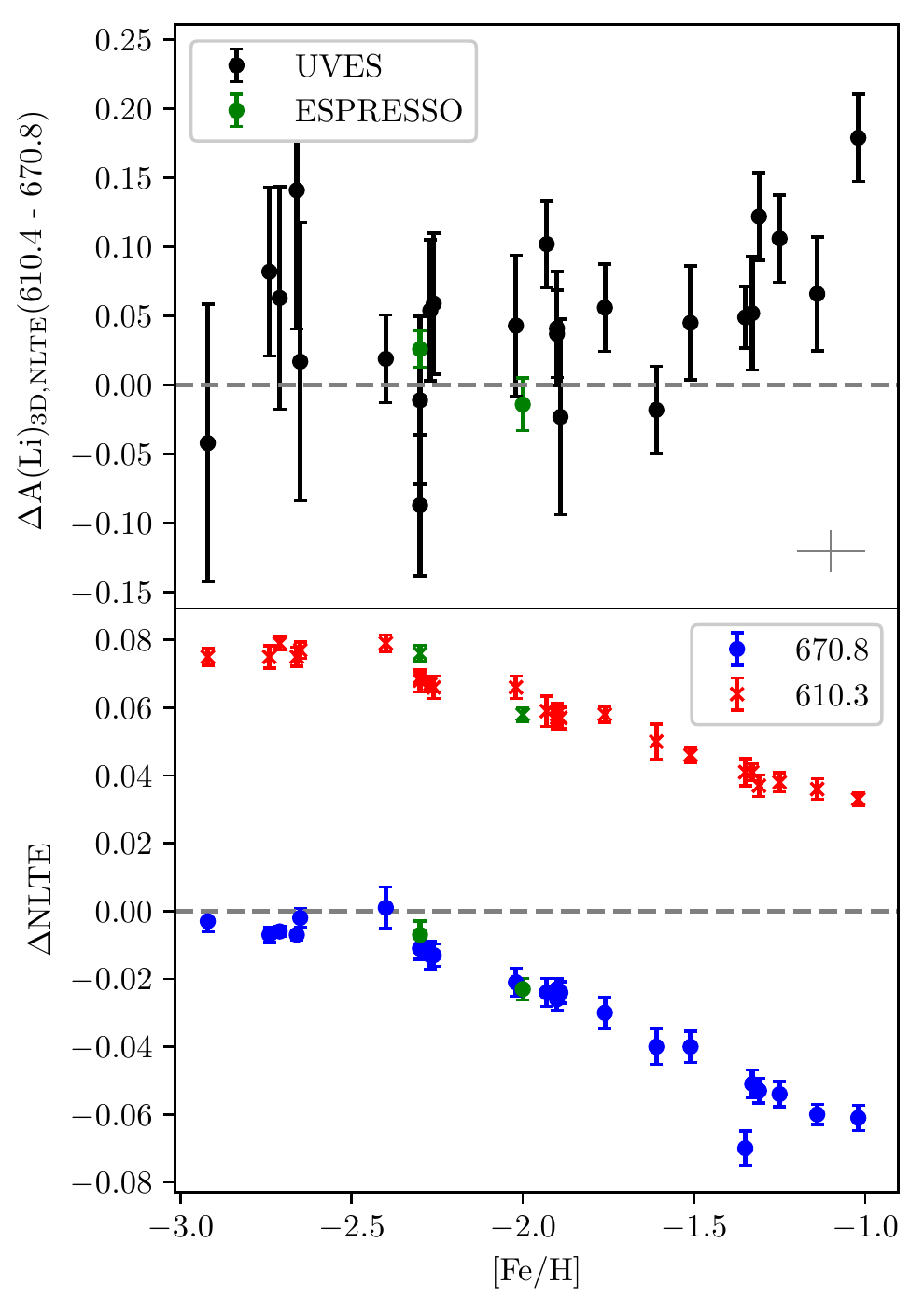}
\caption{Top panel: Comparison between 3D NLTE $\ali$ measurements of the 610.4 and 670.8\,nm lines using UVES and ESPRESSO spectra for a sample of Spite plateau stars (see description in text). A representative systematic error bar due to uncertainties in the stellar parameters is shown in the bottom right corner. Bottom panel: 3D NLTE abundance corrections $\Dn$ (3D NLTE-1D LTE) for the two Li lines.
}
\label{fig:comp_asplund06}
\end{figure}

We use a sample of well-studied MSTO stars on the Spite plateau, with equivalent widths measured from high-quality UVES spectra with $S/N > 400$ \citep{asplund06}. We also analyse two benchmark stars with accurately known literature stellar parameters, HD 140283 \citep{karovicova18} and HD 84937 \citep{casagrande11,vandenberg14,amarsi16b}, and ESPRESSO spectra with $S/N \approx 2000$ (Wang et al., in prep).
We find good agreement between abundances measured from the 610.4\,nm and 670.8\,nm lines as shown in Fig.~\ref{fig:comp_asplund06}, but note that the abundance difference between the two lines is unfortunately rather similar in 3D NLTE and 1D LTE.
A small but significant offset is found at $\met \approx -1.0$, that implies an overestimated abundance of the 610.4\,nm line relative to the 670.8\,nm line.
The average abundance difference between the two lines is $+0.041 \pm 0.055$\,dex in 3D NLTE, compared to $-0.042 \pm 0.053$ in 1D LTE.
We emphasise that the 610.4\,nm feature has an equivalent width of just 0.2\,pm, and the abundance difference could therefore be explained by a minor unrecognised blend with strength of the order 0.05\,pm at $\met = -1$. Curiously, the \ion{Fe}{ii} 610.45\,nm line is located in the blue wing of the \ion{Li}i 610.4\,nm line and has an equivalent width of approximately 0.3\,pm in the solar spectrum. This is in good agreement with the predicted strength \citep[using an oscillator strength from][]{raassen_determination_1998}, and should therefore fall very close to expectations in the metal-poor stars.
While this blend was not included in the spectrum fits by \citet{asplund06} due to its uncertain $\log\,gf$ value, we note that they employed a spectrum fitting method that primarily relied on the strength of the red (unblended) wing of the 610.4\,nm transition.
Assuming the blend did contribute to the measured equivalent widths, the corrected line strengths yield an average abundance difference between the two lines of $+0.022 \pm 0.049$\,dex in 3D NLTE, compared to $-0.062 \pm 0.048$\,dex in 1D LTE.

As our 3D NLTE calculations yield overall lower abundance results than previously, this has important implications for a wide range of science cases. Differences are less significant for e.g. warm and metal-poor stars, where background line opacities in UV are relatively weak. In the measurements presented above, we find for the 22 Spite plateau stars with accurately known temperatures and no obvious Li-enrichment, an average 3D NLTE abundance $\ali = 2.20 \pm 0.05$ based on the 670.8\,nm resonance line, compared to our 1D LTE result $\ali = 2.23 \pm 0.06$ or the 1D NLTE result of \citet{asplund06}, $\ali = 2.21 \pm 0.07$.
This underscores the finding of \citet{asplund03} that the 3D NLTE Li abundances in metal-poor halo stars are quite similar to the 1D NLTE and 1D LTE results due to a fortuitous near cancellation of the 3D and NLTE effects.

\citet{mucciarelli12} performed a commensurate analysis of RGB stars, where their 1D NLTE abundance (with NLTE corrections from \citealt{carlsson94}) measurement $\ali = 0.97 \pm 0.06$ implies an initial abundance in these field stars of $\ali_0 = 2.28$--2.46, where the range represents systematic uncertainties related to the choice of temperature scale and stellar evolution models.
We revert their $\ali$ to 1D LTE using the abundance corrections from \citet{carlsson94}, and then apply our own $\Dn$ as described in Section~\ref{sec:nltec} and Section~\ref{sec:int_rew}; we note that this procedure is not perfect since it depends on the difference in LTE results for the two types of employed 1D model atmospheres (\citealt{mucciarelli12}: Kurucz; here: \atmo) but for our purposes it is sufficiently accurate given the uncertainties in the predicted Li depletion for RGB stars.
Our derived abundances for their stars are significantly lower, at $\ali = 0.81\pm0.06$, thereby implying initial abundances in the range $\ali_0 = 2.12$--2.30, in agreement with the Spite plateau value $\ali = 2.20 \pm 0.05$ derived above. This unexpectedly good agreement could imply that the surface lithium abundance depletion in metal-poor MSTO stars is mainly caused by destruction rather than non-destructive deposition.
We note however that theoretical models of mass-dependent depletion through gravitational settling predict a combination of deposition and destruction \citep{richard05} in good agreement with measurements \citep[e.g.,][]{korn_atomic_2007,lind_atomic_2008,melendez10}. The initial lithium abundance recovered in these cited studies is however still significantly lower than the predicted cosmological abundance level, which opens for additional destruction mechanisms, on the pre-main sequence or otherwise \citep[see][]{fu15}.

\begin{figure}
    \centering
    \includegraphics[width=0.49\textwidth]{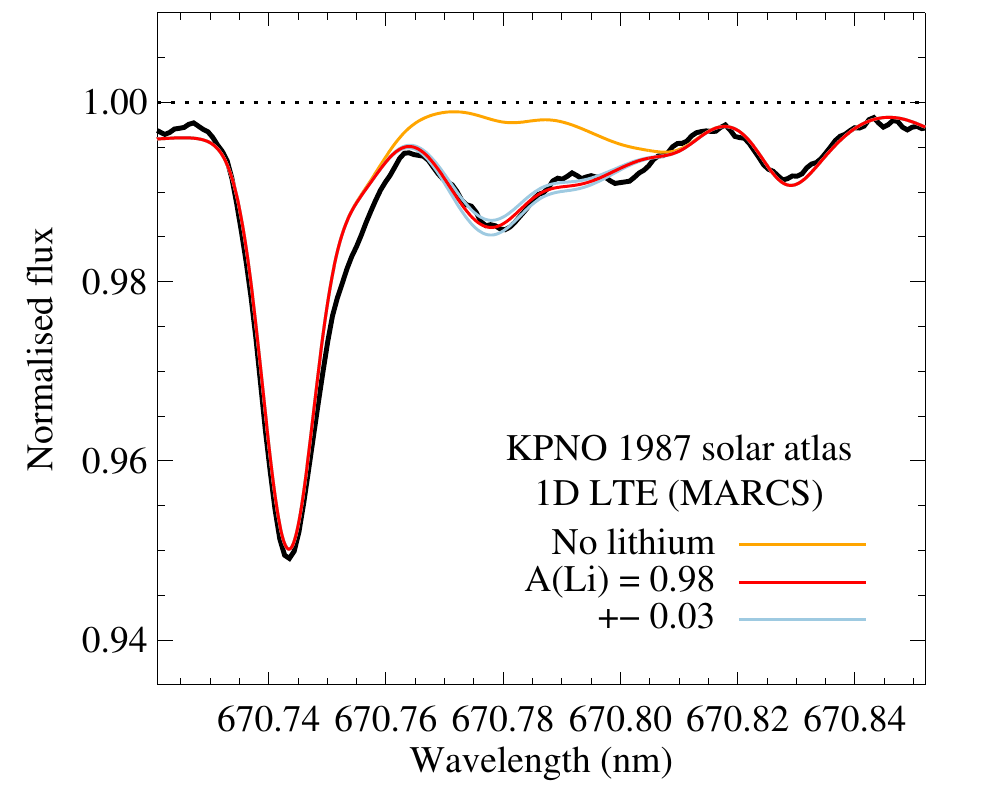}
    \caption{Abundance fit to the solar spectrum in 1D LTE. Our optimum fit has $\textrm{EW} = 0.30\pm0.02$\,pm, which yields a 3D NLTE abundance \bb{$\ali = 0.96 \pm 0.05$} (including systematic errors).}
    \label{fig:solar_fit}
\end{figure}

Finally, we have fitted the \bb{Hamburg solar flux atlas \citep{brault87,neckel_announcement_1999} using the linelist of \citet{ghezzi_measurements_2009} as well as of \citet{melendez_remarkable_2012} in a 1D LTE analysis with MARCS model atmospheres \citep{gustafsson08}. In both cases, our best fit, shown in Fig.~\ref{fig:solar_fit}, the Li line has a line strength of $\textrm{EW} = 0.30\pm0.02$\,pm, which yields a 1D LTE result $\ali = 0.98 \pm 0.03$} (statistical error bars) in excellent agreement with \citet{harutyunyan18}.
This equivalent width corresponds to a 3D NLTE solar abundance of \bb{$\ali = 0.96 \pm 0.05$ (the error bar here includes indicative systematic errors, representing modelling differences comparing 3D/1D and NLTE/LTE, following \citealt{asplund09})},
which is lower than the value $\ali = 1.05 \pm 0.10$ provided in \citet{asplund09}.
It implies that the solar surface Li abundance is depleted by a factor 200 relative to the proto-solar value as implied by the CI meteoritic abundance $\ali = 3.26 \pm 0.05$ (\citealt{lodders09}, renormalised to $\textrm{A(Si)} = 7.51$ from \citealt{amarsi17}).

\section{Conclusions}
\label{sec:con}
In this paper, we presented our 3D NLTE grid, covering the full FGK-type star parameter range, and a broad range in lithium abundances.
We make available our grid alongside a set of interpolation routines in the \breidablik\ package. These interpolation routines can be used to interpolate line profiles to arbitrary stellar parameters using the Kriging technique, achieving a median absolute error of 0.012\,dex; we interpolate line strengths using MLP, achieving a median absolute error of 0.012\,dex when deriving abundance corrections ($\Dn$) and 0.01\,dex when deriving the abundance \bb{from a reduced equivalent width ($\textrm{A}_{\textrm{REW}}$)}. \bb{In addition, 1D departure coefficients are made available through \citep{anish20, anish_mayur_amarsi_2020_3982506}.}

We have shown the importance of taking into account line blocking in the solution of the statistical equilibrium, and find significant differences with respect to previous works.
These lower inferred abundances for all three Li lines
may have important implications for a wide range of science cases. As a demonstration, we redetermine the lithium abundances in samples of metal-poor MSTO and RGB stars, and find that our 3D NLTE abundance estimates are lower than previous 1D NLTE estimates by 0.01 and 0.16\,dex, respectively; after correcting for the predicted effects of dilution due to dredge-up, we find that the lithium abundances measured in metal-poor RGB stars agree well with those measured in the MSTO stars.

In future work, we intend to apply our calculations to varying isotopic ratios in metal-poor stars, in order to revisit the cosmological $^6$Li problem \citep[e.g.][]{smith98,asplund06,lind13}.

\section*{Acknowledgements}
This research was supported by the Australian Research Council Centre of Excellence for All Sky Astrophysics in 3 Dimensions (ASTRO 3D), through project number CE170100013.
This work was supported by computational resources provided by the Australian Government through the National Computational Infrastructure (NCI) under the National Computational Merit Allocation Scheme and the ANU Merit Allocation Scheme \bb{(project y89)}.
MA gratefully acknowledges funding from the Australian Research Council (grants FL110100012 and DP150100250).
AMA acknowledges support from the Swedish Research Council (VR 2016-03765), and the project grant `The New Milky Way' (KAW 2013.0052) from the Knut and Alice Wallenberg Foundation.
KL acknowledges funds from the European Research Council (ERC) under the European Union’s Horizon 2020 research and innovation programme (Grant agreement No. 852977)

\section*{Data Availability}
Our synthetic spectra and interpolation routines are all made publicly available through the \breidablik\ package, available at \url{https://github.com/ellawang44/Breidablik}.

\bibliography{ref}
\bibliographystyle{mnras}

\appendix

\section{Interpolation of the 610.4 and 812.6\,nm lines}
\label{app:rew_other}

In addition to the line strength models for the 670.8\,nm line presented in the main text, we also developed line strength interpolation models for both the 610.4\,nm and 812.6\,nm lines. In Section~\ref{sec:int_rew}, we found that the best performing model is MLP for both $\Dn$ and \bb{$\textrm{A}_{\textrm{REW}}$}, therefore, we use MLP for these new models. In addition, we use the same 5-fold hyperparameter tuning and verification models as previously established to develop these new models.

Table~\ref{table:rew_other_hparams} tabulates the best hyperparameters found through 5-fold cross-validation. The $\Dn$ hyperparameters are similar to the model presented in Section~\ref{sec:mlp} whilst the \bb{$\textrm{A}_{\textrm{REW}}$} hyperparameters differ. This is likely due to the dynamic range and shape of the \bb{$\textrm{A}_{\textrm{REW}}$} data set changing more compared to the $\Dn$ data set when changing Li lines.

\begin{table*}
\begin{tabular}{cc|cccc}
Line strengths & Li line (nm) & $n_l$ & $n$ & $\alpha$ & $f_{\rm{act}}$ \\
\hline
\multirow{2}{5em}{$\Dn$} & 610.4 & 2 & 900 & $10^{-5}$ & ReLU \\
& 812.6 & 2 & 850 & $10^{-1}$ & ReLU \\
\hline
\multirow{2}{5em}{\bb{$\textrm{A}_{\textrm{REW}}$}} & 610.4 & 2 & 750 & $10^{-9}$ & ReLU \\
& 812.6 & 2 & 750 & $10^{-2}$ & ReLU \\
\hline
\end{tabular}
\caption{Optimised MLP hyperparameters for $\Dn$ and \bb{$\textrm{A}_{\textrm{REW}}$} for both the 610.4\,nm and 810.6\,nm line, based on 5-fold cross-validation. $n_l$ is the number of layers, $n$ is the number of neurons per layer, $\alpha$ is the L2 parameter, and $f_{\rm{act}}$ is the activation function.}
\label{table:rew_other_hparams}
\end{table*}

The leave-one-out cross-validation fitting statistics is shown in Table~\ref{table:rew_other}. We find that the errors are smaller than the models presented in Section~\ref{sec:int_rew} because the dynamic range of $\Dn$ and REW for the 610.4\,nm and 812.6\,nm lines are smaller than the dynamic range of the 670.8\,nm line. The training and execution time for these models are comparable to MLP times in Table~\ref{table:line_strengths}.

\begin{table*}
\begin{tabular}{cc|ccccc}
Line strengths & Li line (nm) & MAD & RMS \\
\hline
\multirow{2}{5em}{$\Dn$} & 610.4 & 0.007 & 0.010 \\
& 812.6 & 0.007 & 0.011 \\
\hline
\multirow{2}{5em}{\bb{$\textrm{A}_{\textrm{REW}}$}} & 610.4 & 0.006 & 0.008 \\
& 812.6 & 0.007 & 0.009 \\
\hline
\end{tabular}
\caption{Interpolation errors using a MLP model to predict $\Dn$ and \bb{$\textrm{A}_{\textrm{REW}}$} for both the 610.4\,nm and 810.6\,nm line, based on an average over leave-one-out cross-validation errors using the 3D grid.}
\label{table:rew_other}
\end{table*}

The verification results are shown in Table~\ref{table:ver_other}. Procyon for the 810.6\,nm line is not included as it is immeasurably weak.
Overall, for both $\Dn$ and \bb{$\textrm{A}_{\textrm{REW}}$}, most errors in verification models fluctuate around the MAD leave-one-out cross-validation errors. Therefore, both the $\Dn$ and \bb{$\textrm{A}_{\textrm{REW}}$} final models for the 610.4\,nm and 810.6\,nm line are consistent with their leave-one-out cross-validation errors.

\begin{table*}
\begin{tabular}{ccc|cccc}
Line strengths & Li line (nm) & \diagbox{Method}{Model} & Sun & HD 140283 & Procyon & HD 84937 \\
\hline
\multirow{4}{5em}{$\Dn$} & \multirow{2}{3em}{610.4} & Direct synthesis & 0.021 & 0.069 & 0.080 & 0.057 \\
\cline{3-7}
& & MLP & 0.012 & 0.016 & -0.022 & 0.002 \\
\cline{2-7}
& \multirow{2}{3em}{810.6} & Direct synthesis & 0.001 & 0.060 & & 0.039 \\
\cline{3-7}
& & MLP & 0.009 & 0.015 & & 0.011 \\
\hline
\multirow{4}{5em}{\bb{$\textrm{A}_{\textrm{REW}}$}} & \multirow{2}{3em}{610.4} & Direct synthesis & 1.100 & 2.000 & 1.000 & 2.200 \\
\cline{3-7}
& & MLP & 0.003 & -0.002 & 0.001 & 0.003 \\
\cline{2-7}
& \multirow{2}{3em}{810.6} & Direct synthesis & 1.100 & 2.000 & & 2.200 \\
\cline{3-7}
& & MLP & -0.004 & 0.004 & & 0.009 \\
\hline
\end{tabular}
\caption{Error in predicted $\Dn$ and \bb{$\textrm{A}_{\textrm{REW}}$} using MLP line strength interpolation models trained on the 610.4\,nm and 810.6\, lines on verification models (that were not trained on). The actual $\Dn$ and \bb{$\textrm{A}_{\textrm{REW}}$} are shown in the direct synthesis. The verification models were tailored to specific stars, indicated by name. The 810.6\,nm line is neglected for Procyon as it is immeasurably weak.}
\label{table:ver_other}
\end{table*}

\end{document}